\documentclass[aps,pre,twocolumn,floats,floatfix,showpacs,superscriptaddress]{revtex4}
\usepackage{amsmath,amssymb,eucal,graphicx}
\usepackage[tight]{subfigure}
\renewcommand{\leq}{\leqslant}
\renewcommand{\geq}{\geqslant}

\begin{document}
\title{
Scale-free networks as preasymptotic regimes \\
of superlinear preferential attachment
}

\author{Paul Krapivsky}
\affiliation{Department of Physics, Boston University, Boston, MA
02215, USA}

\author{Dmitri Krioukov}
\affiliation{Cooperative Association for Internet Data Analysis
(CAIDA), University of California, San Diego (UCSD), La Jolla, CA
92093, USA}

\begin{abstract}

We study the following paradox associated with networks growing
according to superlinear preferential attachment: superlinear
preference cannot produce scale-free networks in the thermodynamic
limit, but there are superlinearly growing network models that
perfectly match the structure of some real scale-free networks, such
as the Internet. We obtain an analytic solution,
supported by extensive simulations, for the degree distribution in
superlinearly growing networks with arbitrary average degree, and
confirm that in the true thermodynamic limit, these networks are
indeed degenerate, i.e., almost all nodes have low degrees. We then
show that superlinear growth has vast preasymptotic regimes whose
depths depend both on the average degree in the network and on how
superlinear the preference kernel is. We demonstrate that a
superlinearly growing network model can reproduce, in its
preasymptotic regime, the structure of a real network, if the model
captures some sufficiently strong structural constraints ---
rich-club connectivity, for example. These findings suggest that
real scale-free networks of finite size may exist in preasymptotic
regimes of network evolution processes that lead to degenerate
network formations in the thermodynamic limit.

\end{abstract}


\maketitle

\section{Introduction}

Models of complex networks can be roughly split into two classes:
static and growth models. Static models, such as classical random
graphs~\cite{ErRe59} and their
generalizations~\cite{BoPa03,MaKrFaVa06-phys,SerKriBog07}, generate
a whole network at once, trying to directly reproduce some
properties observed in real network snapshots. Growth models, e.g.,
preferential attachment~\cite{BarAlb99}, construct networks by
adding a node at a time, attempting to provide some insight into the
laws governing network evolution. Compared to static models, it is
generally more difficult to closely match observed network
properties with growth models, because in this case one usually has
less direct control over the properties of modeled networks.

The first growth model that matched the observed Internet topology
surprisingly well, across a wide spectrum of network properties, was
the positive-feedback preference (PFP) model by Zhou and
Mondrag\'on~\cite{ZhoMo04}.
In the model, at each time step, one node is added to the network,
and connected to the existing nodes by two or three links, choosing
different link placement options with different probabilities. The
most important property of the model is that the probability to
connect a new node to the existing nodes of degree $k$ is
a superlinear function of $k$.
Although there are many other models of
the Internet evolution, e.g.~\cite{FaKra03,SeBoDi05,SeBoDi06},
the PFP model stands apart as it gives rise to the following
unresolved paradox. On the one hand the model matches perfectly the
observed Internet, while on the other hand, since it is explicitly based on
preferential attachment with a superlinear preference kernel, it cannot produce,
in the thermodynamic limit, any scale-free networks~\cite{KraReLe00}, including the Internet.

Here we resolve this paradox by showing that superlinear
preferential attachment can have vast preasymptotic regimes.
Specifically, we first find an analytic asymptotic solution for
superlinearly growing networks with arbitrary average degree,
confirming that the asymptotic regime is indeed degenerate ---
regardless of the average degree, only a finite number of nodes
have high degrees (Section~\ref{sec:analytic-solution}). However, in
Section~\ref{sec:preasymptotic}, we show that if the preference
kernel is not too superlinear and if the average degree is not too
low, then this asymptotic regime becomes noticeable only at network
sizes that are orders of magnitudes larger than the size of any real
network, including the Internet. We thus half-resolve the paradox by
showing that the PFP model {\em can}, in fact, match the Internet.
Section~\ref{seq:rcc-vs-jdd} resolves the other half, by explaining
why the model {\em does} so: its design implicitly reproduces the degree
correlations in the Internet, which are known to define almost all
important topological properties, except
clustering~\cite{MaKrFaVa06-phys,SerKriBog07}. We conclude in
Section~\ref{sec:conclusion} with an outline of our findings and
their implications.

\section{Asymptotic degree distribution}
\label{sec:analytic-solution}

In this section we derive the analytic solution for the degree
distribution of superlinearly growing networks (SLGNs) in the
thermodynamic limit. We begin by recalling what is known for
networks grown by adding a single link per node (the average degree
$\bar{k}\approx2$), and then generalize to the case with multiple
links.

\subsection{Single link per node}

The case when a new node attaches to exactly one existing target (or
host) node is well-studied~\cite{KraReLe00,KraRe01,OlSp05}. Let the
probability that the new node selects a host node of degree $k$ be
\begin{equation}
\label{attach}
k^\delta/\sum_{j=1}^N (k_j)^\delta,
\end{equation}
where the summation is over all $N$ existing nodes and $k_j$'s are
their degrees. Then the asymptotic degree distribution  is a stretched
exponential for sub-linearly growing networks ($\delta<1$) and a power law
for linearly growing networks ($\delta=1$). Superlinearly growing networks
with $\delta>1$ are asymptotically star graphs.

Specifically, if $\delta>2$, then the number of nodes with degrees
$k>1$ remains finite in the thermodynamic limit $N\to\infty$,
meaning that almost all nodes have degree $1$, $N_1(N) \approx N$.
If $3/2<\delta<2$, then the number $N_2(N)$ of nodes with degree $2$
(degree-$2$ nodes) grows as $N^{2-\delta}$, while the number of
nodes with degrees $k>2$ remains finite. If $4/3<\delta<3/2$, then
$N_3(N) \sim N^{3-2\delta}$, and the number of nodes with degrees
$k>3$ is finite. In other words, there is an infinite series of
``phase transitions'' at critical values $\delta_p=1+1/p$, where
$p=1,2,3,\ldots$, and the degree distribution in SLGNs with $\delta$
lying between these critical values, $\delta_p<\delta<\delta_{p-1}$
($\delta_0\equiv\infty$), is given by
\begin{equation}
\label{eq:Nk,m1} N_{k}/N \sim
\begin{cases}
N^{(k-1)(1-\delta)} & \text{if $1 \leq k \leq p$;}\\
1/N                 & \text{otherwise.}
\end{cases}
\end{equation}

In what follows we also consider the {\em extremal growth\/}
rule, which formally corresponds to the $\delta\to\infty$ limit, and
specifies that a new node attaches to the existing node with the
maximum degree. If there are several nodes with the same maximum
degree, then the host node is randomly selected among them.
SLGNs grown according to this rule stay stars throughout
their evolution, assuming they are stars at the beginning. If
an SLGN is not initially a star, then extremal growth evolves it
to almost a star, with all new nodes attaching to a maximum-degree
node in the initial graph.

Adding one link per node results in growing trees, which are not
good models of real complex networks that all have strong
clustering. But even if we are not concerned with the models' realism,
there is another reason to consider SLGNs with multiple links added
per new node.

While for sub-linearly growing networks, adding more than one link
should not qualitatively change the degree distribution, this
modification may have a more prominent effect on the degree
distributions in superlinearly growing networks. Indeed, the more
links per node we add in SLGNs with a finite $\delta$, the stronger
the deviations from stars we obviously expect to observe. In view of the
PFP model paradox, one might even start suspecting that multiple
links may resurrect power laws. We thus have to exercise more care
dealing with multiple-link SLGNs. In what follows, we first consider
them under the extremal growth rule, and then remove this
restriction.

\subsection{Multiple links per node. Extremal growth.}

We denote by $m$ the number of links added per new node. The
PFP model uses a superposition of the $m=2$ and $m=3$ cases, and a
combination of the   following two link placement options: place a
link either between the new and host nodes, or between the host and
another existing node, called the peer node.
Links are always placed such that the subgraph induced by the new links
is connected and contains the new node, so that the network stays connected
at each time step.
For concreteness, we shall
assume that $m$ is a fixed positive integer, and consider cases with
different $m$ separately. Another important restriction
is that we construct simple graphs, i.e., self-loops and multiple
links between the same two nodes are not allowed.

\begin{figure}[tb]
    \centering
    \subfigure[Two hosts]
        {\includegraphics[width=.45\linewidth]{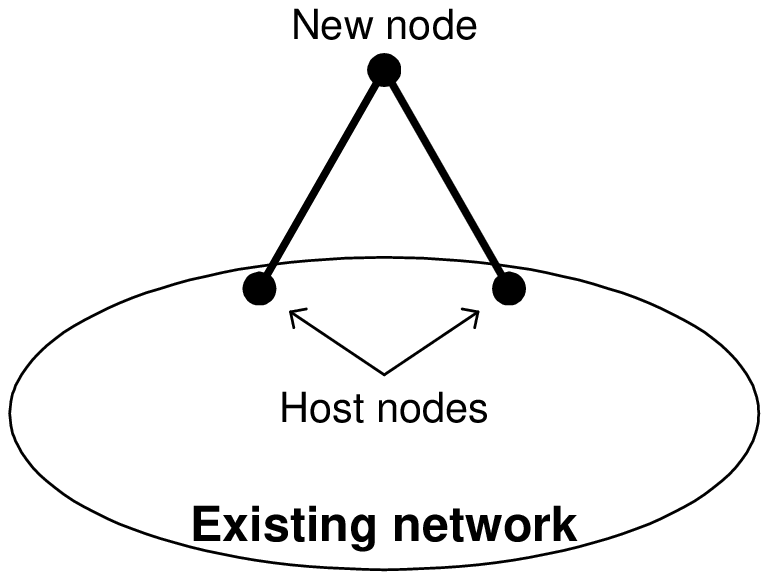}
        \label{fig:link-placement-a}}
    \subfigure[One host]
        {\includegraphics[width=.45\linewidth]{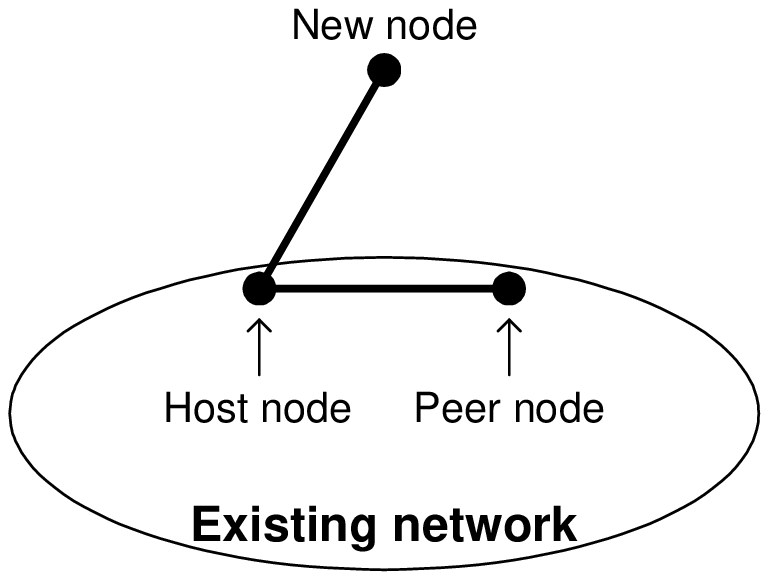}
        \label{fig:link-placement-b}}
    \caption{Link placement options for two links.}
    \label{fig:link-placement}
\end{figure}

We first focus on the case with $m=2$. In this case we have only two
options to place two links (see Fig.~\ref{fig:link-placement}):
place both links between the new and host nodes, or place one link
between the new and host nodes, and place another link between the
host and peer nodes. The both options, or any their superposition,
produce the same result. Let the initial network be two disconnected
nodes. Adding the third node according to the extremal growth rule
creates the star graph with three nodes. We shall represent our
graphs by their degree sequences $(k_1,\ldots,k_N)$. The degree
sequence representation turns out to define, up to an isomorphism,
the graphs grown according to our extremal growth rule. The star
graph after the first step is $(2,1,1)$ in this representation.
Applying the extremal growth rule to add the fourth node, we obtain $(3,2,2,1)$, and
then $(4,3,2,2,1)$, $(5,4,2,2,2,1)$, and generally
\begin{equation}
\label{book}
(N-1,N-2,\,\underbrace{2,\ldots,2}_{N-3}\,,1)
\end{equation}

We prove (\ref{book}) by induction. We already checked its validity
for small $N$. Assuming that (\ref{book}) holds for some $N>4$, we
establish it for $N+1$. If we place the two links according to the
first option, shown in Fig.~\ref{fig:link-placement-a}, then the new
node attaches to the nodes with degrees $N-1$ and $N-2$. Thus degree
$N-1 \mapsto N$ and $N-2 \mapsto N-1$, the new node acquires degree
$2$, other degrees do not change, and the new graph has indeed the
same structure (\ref{book}). If we use the other link placement
option shown in Fig.~\ref{fig:link-placement-b}, we must choose the
node of degree $N-2$ as the host node. We cannot attach the new node
to the node of degree $N-1$ because this latter node is already
connected to all other nodes, and therefore we cannot add the second
link between this hub node and any peer node. Selecting the node of
degree $N-2$ as the host, we notice that it is connected to all
other nodes, except the degree-$1$ node. Therefore this latter node
is the only choice for the peer node. Hence $N-2\mapsto N$ and
$1\mapsto 2$, the new node acquires degree $1$, and other degrees do
not change. Thus the new graph has the same structure (\ref{book}).

\begin{figure}[tb]
     \includegraphics[width=0.45\linewidth]{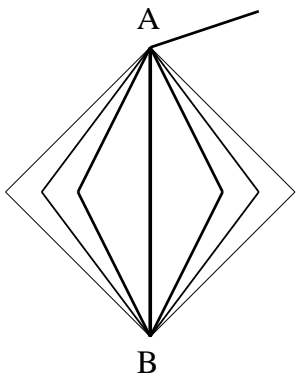}
  \caption{Open $2$-book (8,7,2,2,2,2,2,2,1).}
\label{open}
\end{figure}

We shall call the graphs series (\ref{book}) the {\em open
$2$-books}. The justification for this name is as follows. The link
between the two nodes of highest degrees, denoted by $A$ and $B$, is
the {\em binding} of an open book. Each degree-$2$ node is connected
to $A$ and $B$, and the resulting triangle is a {\em page}. Thus, an
open $2$-book contains $N-3$ triangular pages. Finally, the link
between the highest-degree node $A$ and the dangling degree-$1$ node
is a built-in {\em bookmark}. The open $2$-book graph with $N=9$
nodes is shown in Fig.~\ref{open}.

Note that we can call a star an open $1$-book. It does not have
bookmarks, its binding is the hub node, and its $N-1$ pages are all
the links.

\begin{figure}[tb]
    \centering
    \subfigure[Three hosts]
        {\includegraphics[width=.45\linewidth]{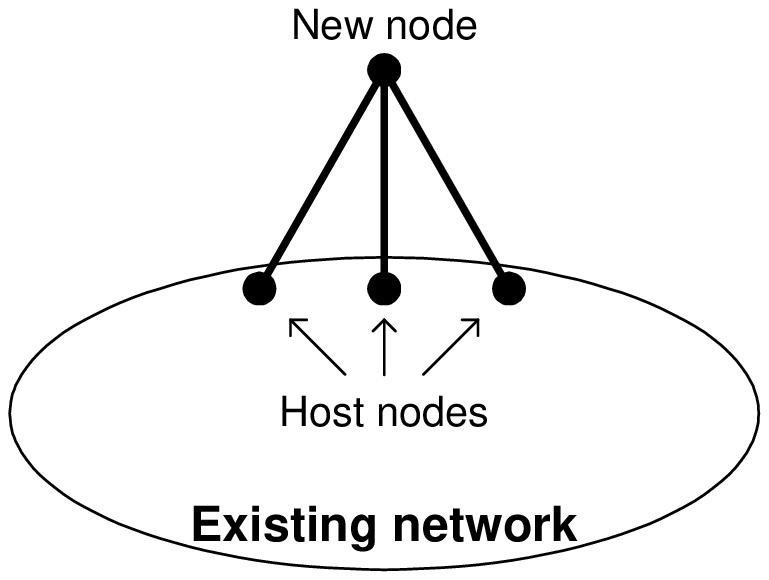}
        \label{fig:link-placement-m=3a}}
    \subfigure[Two hosts]
        {\includegraphics[width=.45\linewidth]{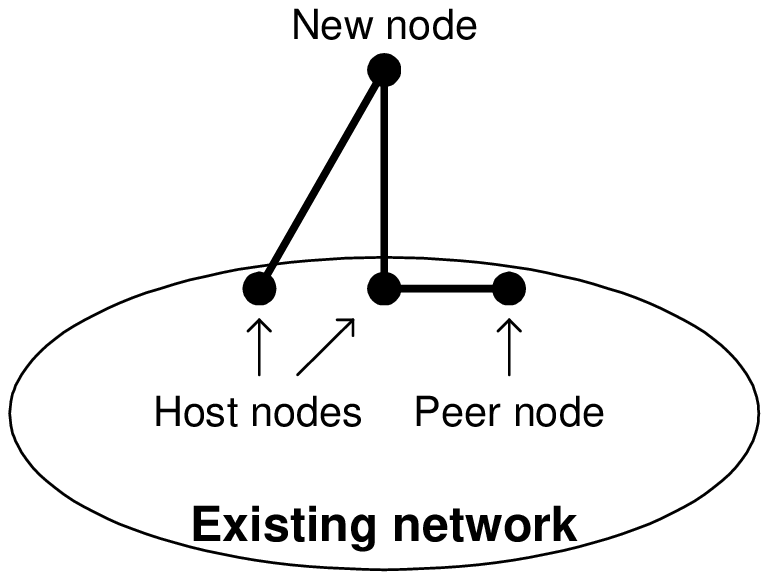}
        \label{fig:link-placement-m=3b}}\\
    \subfigure[One host]
        {\includegraphics[width=.45\linewidth]{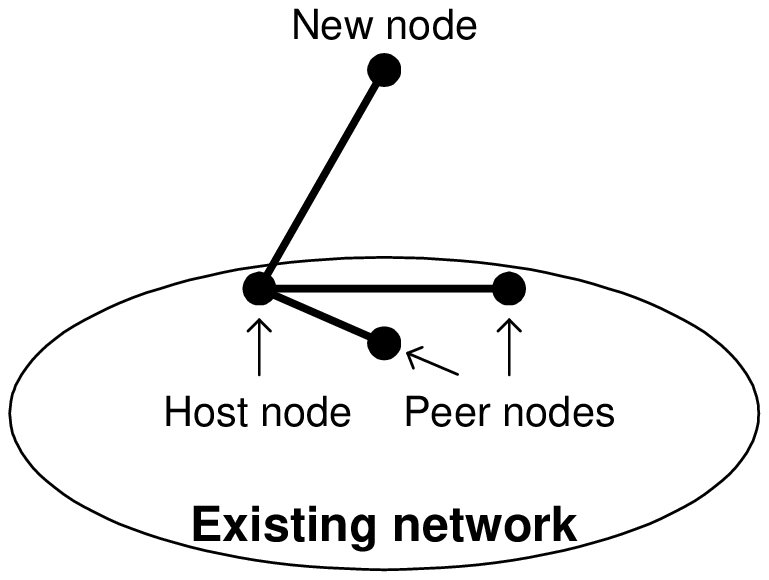}
        \label{fig:link-placement-m=3c}}
    \subfigure[One host]
        {\includegraphics[width=.45\linewidth]{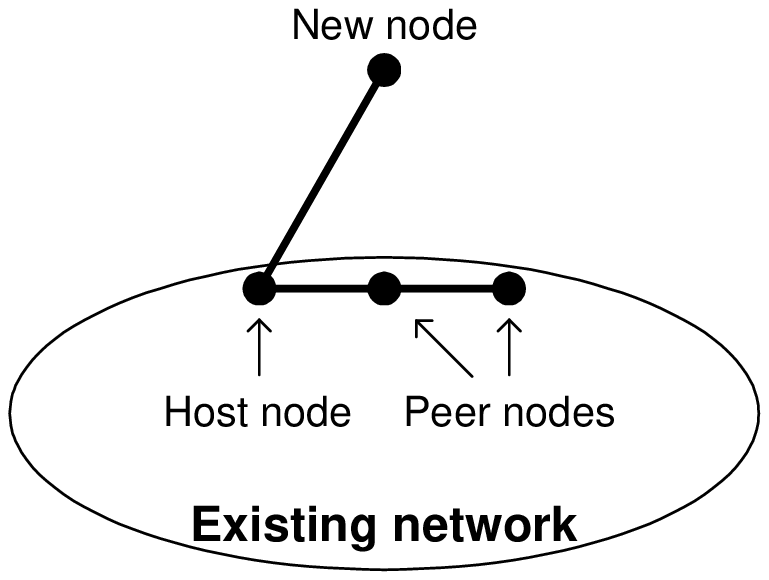}
        \label{fig:link-placement-m=3d}}
    \caption{Link placement options for three links.}
    \label{fig:link-placement-m=3}
\end{figure}

We now move to the case with three links added per new node, $m=3$.
Generalizing the
link placement options for two links, there are four options
for placing three links, shown in Fig.~\ref{fig:link-placement-m=3}.
Choosing the first option with three host nodes
(Fig.~\ref{fig:link-placement-m=3a}), the application of the
extremal growth rule to the initial graph $(0,0,0)$ yields the graph
series $(3,1,1,1)$, $(4,3,2,2,1)$, $(5,4,3,3,2,1)$,
$(6,5,4,3,3,2,1)$, $(7,6,5,3,3,3,2,1)$, and generally
\begin{equation}
\label{book2}
(N-1,N-2,N-3,\,\underbrace{3,\ldots,3}_{N-5}\,,2,1)
\end{equation}
Using the same logic as in the $m=2$ case, one can prove that
the extremal growth indeed produces (\ref{book2}). The link
placement option in Fig.~\ref{fig:link-placement-m=3c} results in
exactly the same graph series. Placing links as in
Fig.~\ref{fig:link-placement-m=3d}, we obtain almost the same graph
series, except that the first graph is $(2,2,1,1)$. The option in
Fig.~\ref{fig:link-placement-m=3b} leads to a different graph
series, but almost all nodes still have degree $3$.

We call the graph series (\ref{book2}) {\em open
$3$-books}. The binding is the triangle $ABC$
connecting the three nodes $A$, $B$, and $C$ of highest degrees
$N-1$, $N-2$, and $N-3$. Each page is a tetrahedron $ABCD$, where
$D$ is one of the $N-5$ degree-$3$ nodes. Thus, an open $3$-book has
$N-5$ tetrahedral pages. It also has two bookmarks: triangle $ABE$
and link $AF$, where $E$ and $F$ are the nodes of degrees $2$ and~$1$.

Generalizing to an arbitrary $m$, we notice that there are
combinatorially many possibilities to place $m$ links. In general,
they lead to different graph series. For concreteness, in the rest
of this paper we focus on the simplest option with no peer nodes and
$m$ hosts, i.e., the generalization of
Figs.~\ref{fig:link-placement-a},\ref{fig:link-placement-m=3a}. In
this case, if $N$ is sufficiently large, i.e., $N>2m$, then the
resulting graphs are
\begin{equation}
\label{bookm}
(N-1,\ldots,N-m,\,\underbrace{m,\ldots,m}_{N-2m+1}\,,m-1,\ldots,1)
\end{equation}
These graphs are {\em open $m$-books}. If we imagine them placed in
an $m+1$-dimensional ambient space, then they contain:
\begin{itemize}
\item one $2$-codimensional binding, i.e., the $m-1$-simplex $A_1\ldots
A_m$ composed of the $m$ highest-degree nodes;
\item $N-2m+1$ $1$-codimensional pages, i.e., $m$-simplices
$A_1\ldots A_m D$, where $D$ is one of the $N-2m+1$ degree-$m$
nodes;
\item $m-1$ bookmarks of codimensions $2, 3, \ldots, m$, i.e.,
$m-1$-, $m-2$-, \ldots, and $1$-simplices (one simplex of each
dimension), composed of links interconnecting highest-degree nodes
and nodes of degrees $k<m$.
\end{itemize}

The notion of an open book appears in mathematics~\cite{Giroux05},
where it finds various applications, e.g., as a tool to establish
connections between contact geometry and topology. In its simplest
definition, an open book is a fibration of a manifold by a
collection of $1$-codimensional submanifolds (pages), joined along a
$2$-codimensional submanifold (binding). Open books with bookmarks
(formal definitions are obvious) seem natural, and perhaps they will
find applications, too.

\subsection{Removing the extremal growth restriction}

In this section we outline the logic behind removing the extremal
growth condition. (A more detailed exposition is presented in
the Appendix.) We assume that $\delta$ has a finite
value. First, we estimate the probability that an SLGN remains an
open book. We then characterize the deviations from the open book
structure. For clarity, we consider the simplest case with $m=2$ and
the first link placement option in Fig.~\ref{fig:link-placement-a}.

Consider a network of large size $j$, so that $j \approx j-1 \approx
j-2 \approx j-3$, and suppose that it is an open
$2$-book~(\ref{book}). Avoiding multiple links between the same pair
of nodes, the probability $\mathcal{P}_{j \mapsto j+1}$ that after
adding one node the network preserves its open book structure is
approximately
\begin{equation}
\label{jj}
\mathcal{P}_{j\mapsto j+1}\approx
\frac{j^\delta+j^\delta}{j^\delta+j^\delta+j\cdot 2^\delta}\cdot
\frac{j^\delta}{j^\delta+j\cdot 2^\delta}\,.
\end{equation}
Indeed, the first factor is the probability that one of the two
nodes of degree $\approx j$ is selected as the first host, while the
second factor is the probability that the other such node is
selected as the second host, in which case the network preserves its
approximate open book structure. Using (\ref{jj}) we estimate the
probability $\mathcal{P}_N$ that upon reaching size $N$ the network
is still an open $2$-book
\begin{eqnarray}
\label{PN}
\mathcal{P}_N
&\approx&\prod^N_{j=2} \frac{1}{1+\left(\frac{2}{j}\right)^{\delta-1}}
\cdot\frac{1}{1+2\left(\frac{2}{j}\right)^{\delta-1}}\nonumber\\
&\sim&
\begin{cases}
\text{finite in the limit $N\to\infty$}&\text{if $\delta>2$};\\
N^{-6}&\text{if $\delta=2$};\\
e^{-a\,N^{2-\delta}}, \quad a=\frac{3 \cdot 2^{\delta-1}}{2-\delta} & \text{if $\delta<2$}.
\end{cases}
\end{eqnarray}
We thus see that if $\delta$ is sufficiently large, viz.\
$\delta>2$, then there is a finite probability that the network
preserves its open $2$-book structure throughout the entire
evolution. This observation implies that even if it is not an open
book, the distortion of the open book structure is finite, e.g., a
finite number of nodes have degree $k>2$, degrees of nodes $A$ and
$B$ in Fig.~\ref{open} are respectively lower, etc.

However, if $\delta\leq 2$, the network is not an open book with
high probability. But even though the exact open book structure is
almost surely destroyed, the distortion is still asymptotically
small and admits analytic estimates. Indeed, let us first estimate
the number $N_3(N)$ of degree-$3$ nodes in an $N$-sized SLGN with
$m=2$ and $\delta\leq 2$. This number grows if instead of connecting
to the highest-degree node with probability $\mathcal{P}_{N\mapsto
N+1}$ in Eq.~(\ref{jj}), the new node selects the other option and
connects to a degree-$2$ node with probability
$1-\mathcal{P}_{N\mapsto N+1}$. Therefore
\begin{equation}
\label{N3}
\frac{d N_3}{d N} \approx 1-\mathcal{P}_{N\mapsto N+1}
\approx 3\left(\frac{2}{N}\right)^{\delta-1},
\end{equation}
where we have neglected loss terms describing the decrease of the
number of degree-$3$ nodes due to new nodes connecting to them and
changing their degrees to $4$ or $5$. These loss terms, as well as
corrections to the approximate expression for $\mathcal{P}_{N\mapsto
N+1}$ in \eqref{jj}, are sub-leading, as we show in Appendix.
The integration of Eq.~(\ref{N3}) gives
\begin{equation}
\label{N3-sol}
N_3(N)\approx
\begin{cases}
6\ln N & \text{if $\delta=2$};\\
a\,N^{2-\delta} & \text{if $\delta<2$},
\end{cases}
\end{equation}
which we juxtapose against simulations in
Section~\ref{sec:preasymptotic}.  We thus see that the number of
degree-$3$ nodes grows sublinearly with $N$, and consequently their
proportion in the thermodynamic limit is infinitesimal. We also note
that the solution in Eq.~(\ref{N3-sol}) allows us to compactly
rewrite Eq.~(\ref{PN}) as
\begin{equation}
\mathcal{P}_N \sim e^{-N_3(N)}.
\end{equation}

The obvious generalization of (\ref{N3}) for higher degrees is
\begin{equation}
\label{Nk} \frac{dN_k}{dN}\sim \frac{N_{k-1}}{N^\delta}.
\end{equation}
Solving recursively yields the connectivity transitions quite
similar to those in the $m=1$ case (\ref{eq:Nk,m1})
\begin{equation}
\label{eq:Nk,m2} N_{k}/N\sim
\begin{cases}
N^{(k-2)(1-\delta)} & \text{if $2 \leq k \leq p+1$};\\
1/N                 & \text{otherwise},
\end{cases}
\end{equation}
for any $\delta$ such that $\delta_p < \delta < \delta_{p-1}$, where
$\delta_p=1+1/p$ and $p=1,2,3,\ldots$. The only difference between
the degree distributions for the $m=1$ and $m=2$ cases
(Eqs.~(\ref{eq:Nk,m1}) and~(\ref{eq:Nk,m2})) is that the latter is
the former shifted along the $k$-axis to the right by $1$
(Eq.~(\ref{eq:Nk,m2}) is Eq.~(\ref{eq:Nk,m1}) with $k \mapsto k-1$).

Therefore, the same infinite series of connectivity transitions
appear for any $m \geq 1$, and the asymptotic degree distribution is
given by
\begin{equation}
N_{k}/N\sim
\begin{cases}
N^{(k-m)(1-\delta)} & \text{if $m \leq k \leq p+m-1$};\\
1/N                 & \text{otherwise}.
\end{cases}
\end{equation}

\section{Preasymptotic regime}
\label{sec:preasymptotic}

We have shown that all SLGNs are asymptotically open books, while
Zhou and Mondrag\'on~\cite{ZhoMo04} showed that a specific SLGN of a
finite size exhibited clean power laws. Another apparent
disagreement is that according to our analysis, $N_k/N \to 0$ for all
$k>m$, while the PFP model simulations show that $N_k
\sim N$ for all $k$. The explanation of these paradoxes lies in
the fact that the PFP model has a vast preasymptotic regime, and
both the Internet size and sizes achievable in simulations lie deep
within this regime. In this section, we describe two main factors that
render this regime extremely vast for the PFP-modeled Internet.

The first factor is that $\delta$ in the PFP model exceeds $1$ only
slightly (specifically, $\delta\approx1.15$ in \cite{ZhoMo04}). For
clarity, let us focus on the following concrete example. The
proportion of degree-$3$ nodes $N_3/N$ in the $m=2$ case scales as
$N^{1-\delta}$, so that if $\delta$ is close enough to $1$, then the
deviation of $N_3(N)$ from the linear growth may be hard to observe
for insufficiently large $N$. Indeed, $N=10^4$ (the order of the
Internet size) and $\delta=1.15$ substituted in Eq.~(\ref{N3-sol})
yield $N_3/N\approx 0.98$, contradicting the assumption
made to derive Eq.~(\ref{N3-sol})
that the network is almost an open $2$-book and hence
$N_3/N \ll 1$. This contradiction means that we are very far from the
asymptotic regime. Even if we choose $N=10^{10}$ (almost two
autonomous systems per person!), the ratio $N_3/N$ goes down only to
$12\%$, so it is still far from negligible. To get it down to $1\%$,
we would need $N = 10^{17}$, non-achievable in simulations.

The second factor deepening the preasymptotic regime is $m>1$.
The larger $m$, the slower the decay of
$N_k/N \sim N^{(k-m)(1-\delta)}$ for $k>m$, the deeper the
preasymptotic regime. For example, using the results from
\cite{KraRe01} for $N_3(N)$ in the $m=1$ case, we find that the
Internet size $N=10^4$ and $\delta=1.15$ yield $N_3/N \approx 0.24$,
and to get it down to $1\%$, we would need only $N=10^8$, while
$N=10^{10}$ makes it $0.4\%$---all the numbers are substantially
lower than in the $m=2$ case.

\begin{figure}[tb]
    \centering
    \subfigure{\includegraphics[height=1.36in]{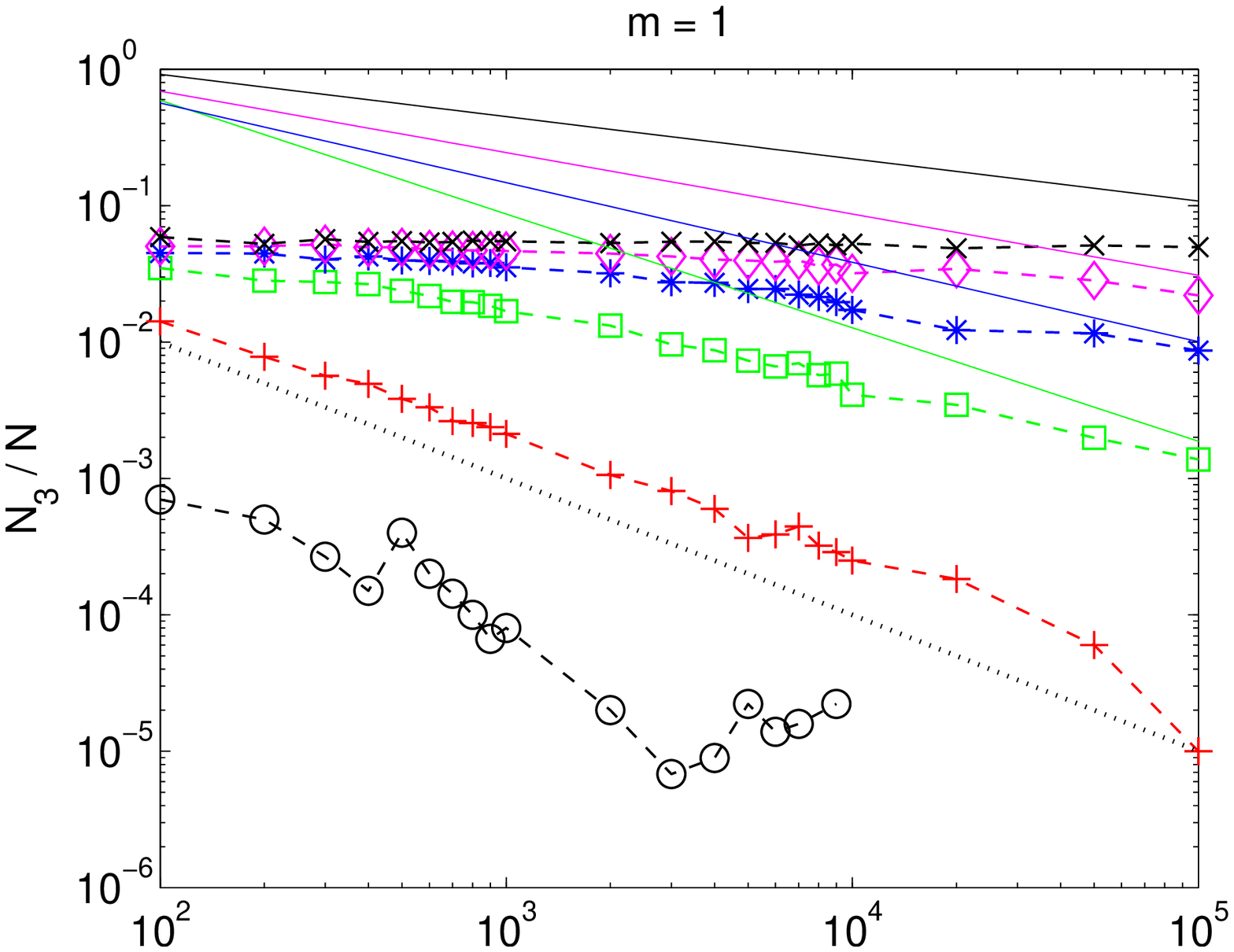}}\hfill
    \subfigure{\includegraphics[height=1.36in]{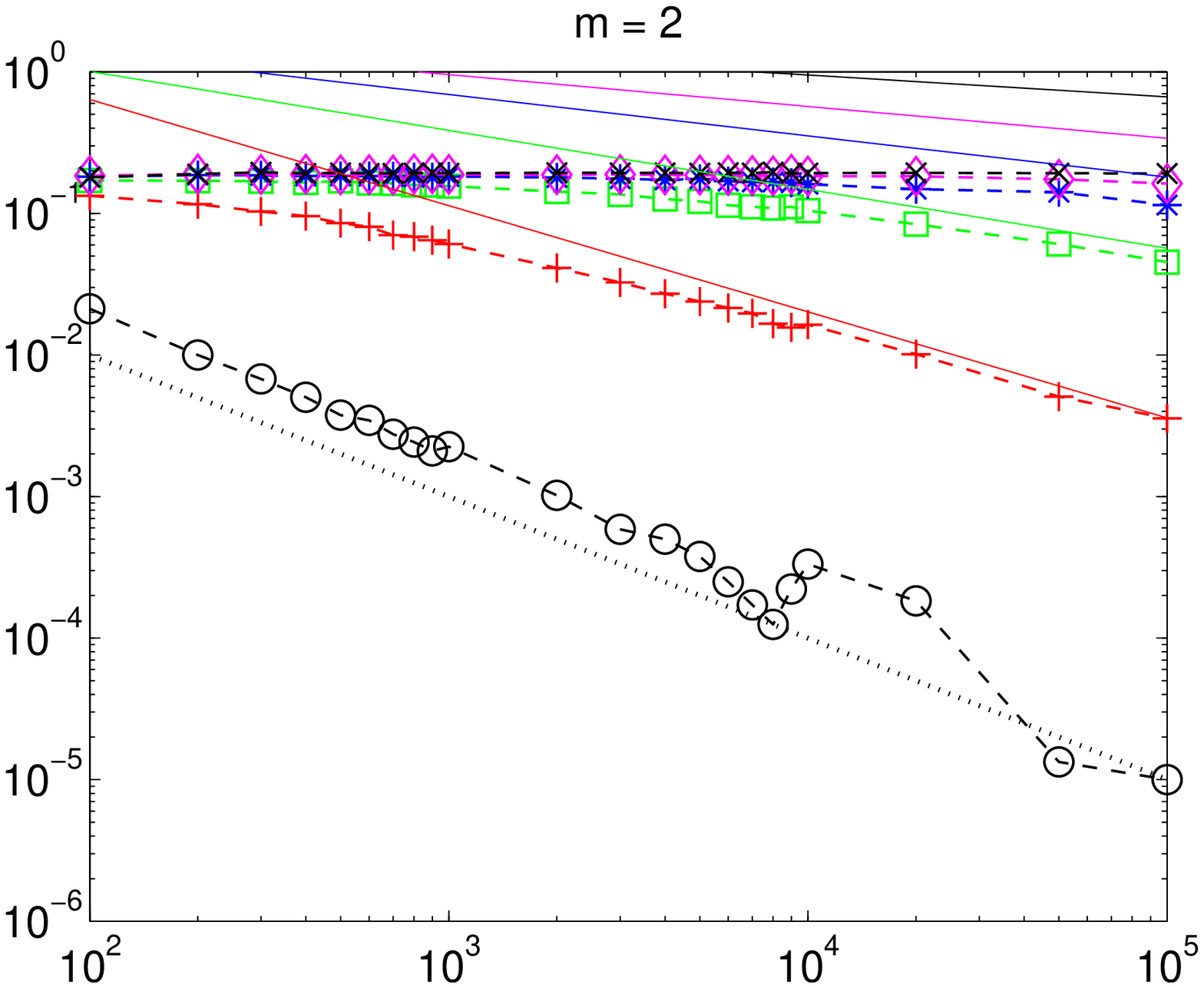}}\\
    \subfigure{\includegraphics[height=1.44in]{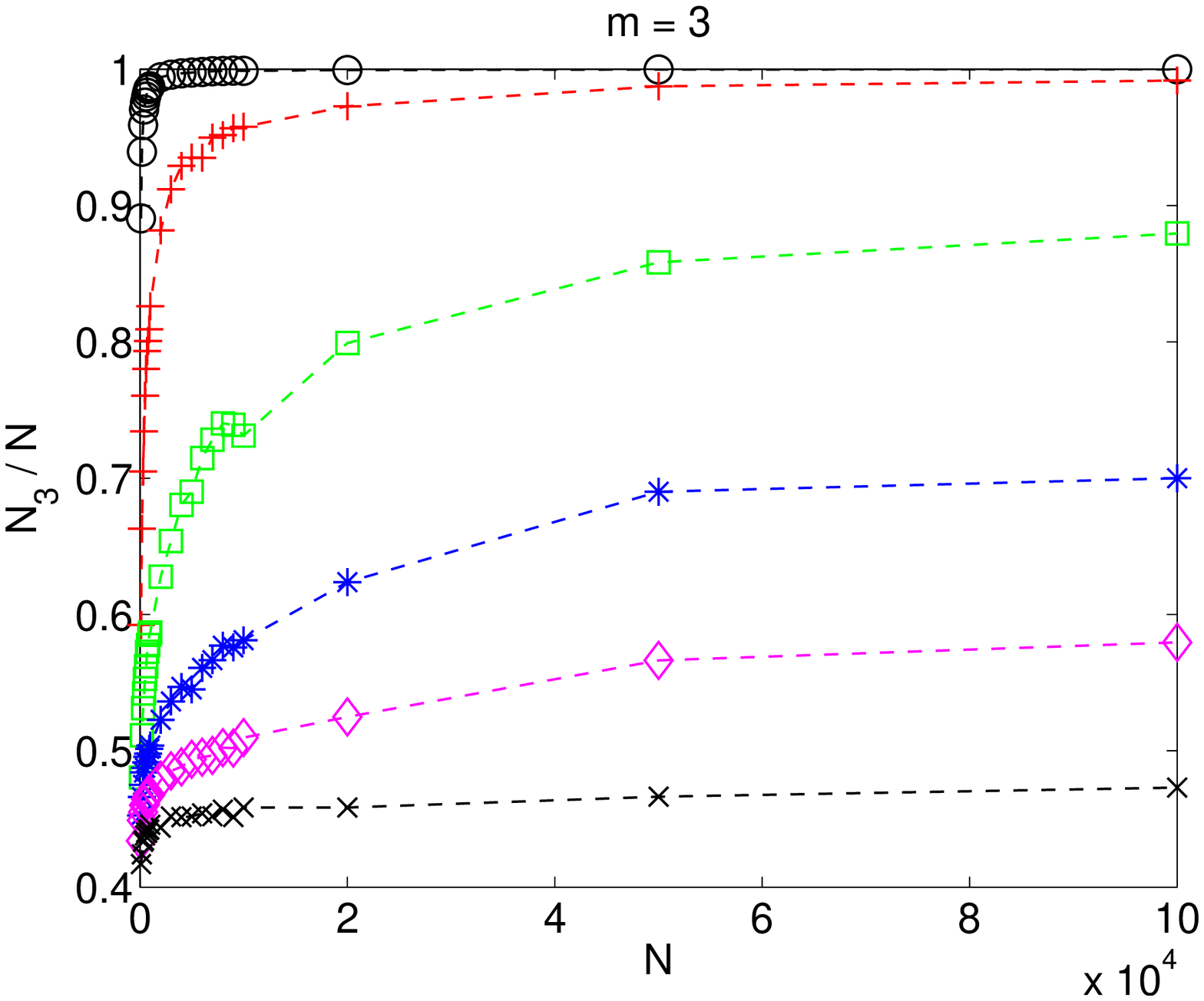}}\hfill
    \subfigure{\includegraphics[height=1.44in]{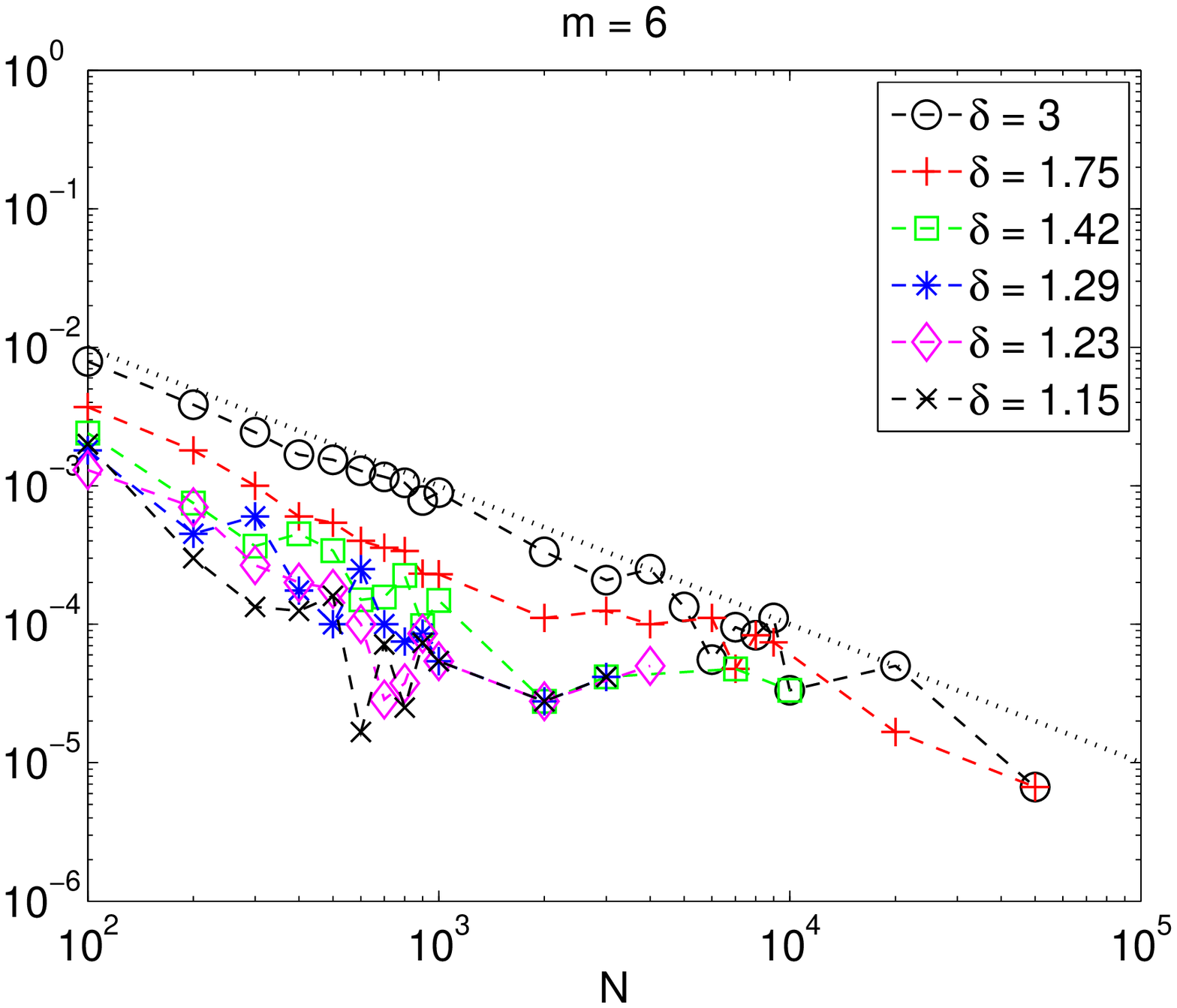}}
    \caption{(Color online) Scaling of the proportion of degree-$3$ nodes $N_3/N$
    in SLGNs with different $\delta$ and $m$. The solid lines are
    the analytic predictions for the leading term from \cite{KraRe01}
    for $m=1$ and Eq.~(\ref{N3-sol}) for $m=2$. The dashed lines are
    simulations. The dotted line is $1/N$.}
    \label{fig:N3/N}
\end{figure}

\begin{figure*}[tb]
    \centering
    \subfigure{\includegraphics[height=1.41in]{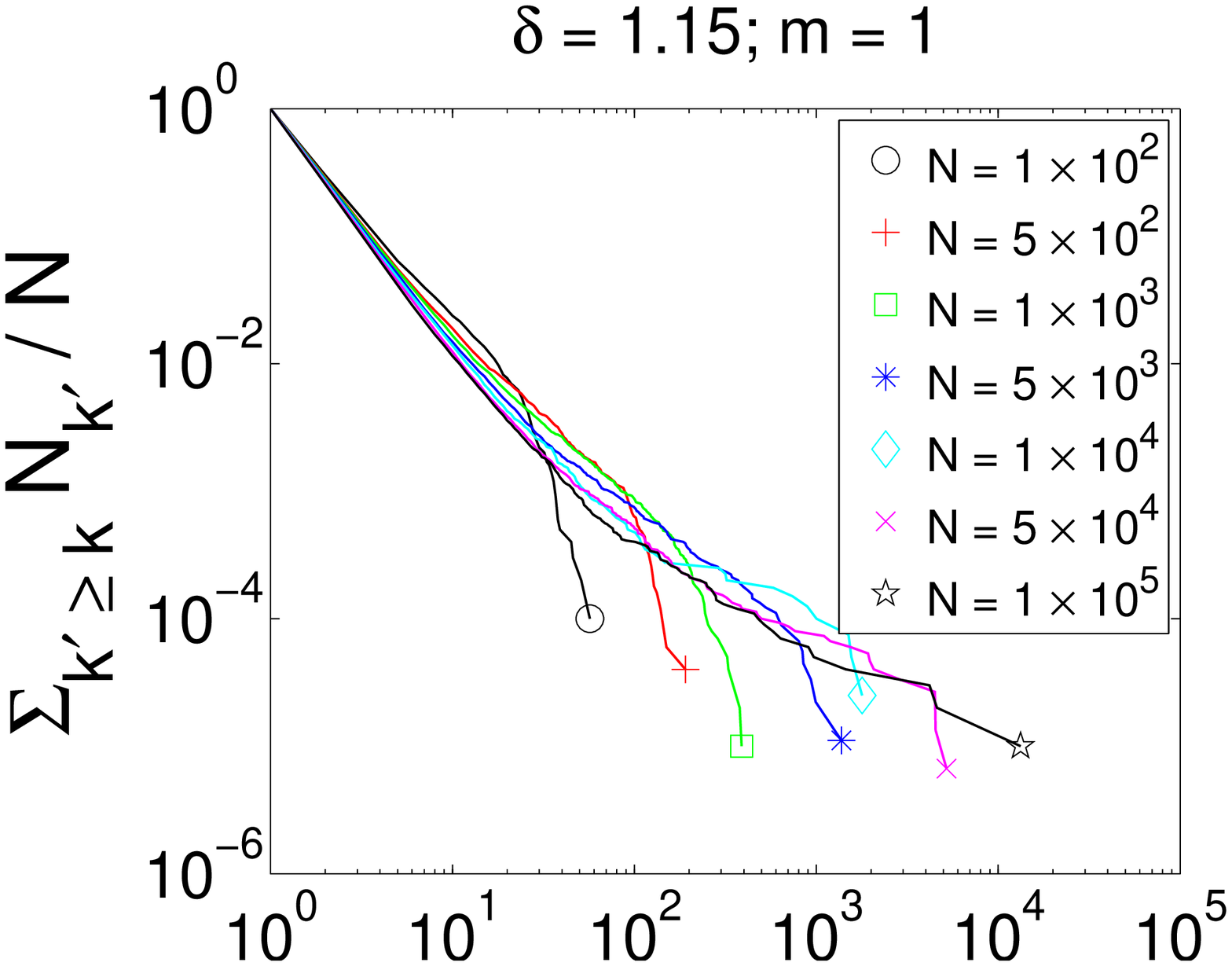}}\hfill
    \subfigure{\includegraphics[height=1.41in]{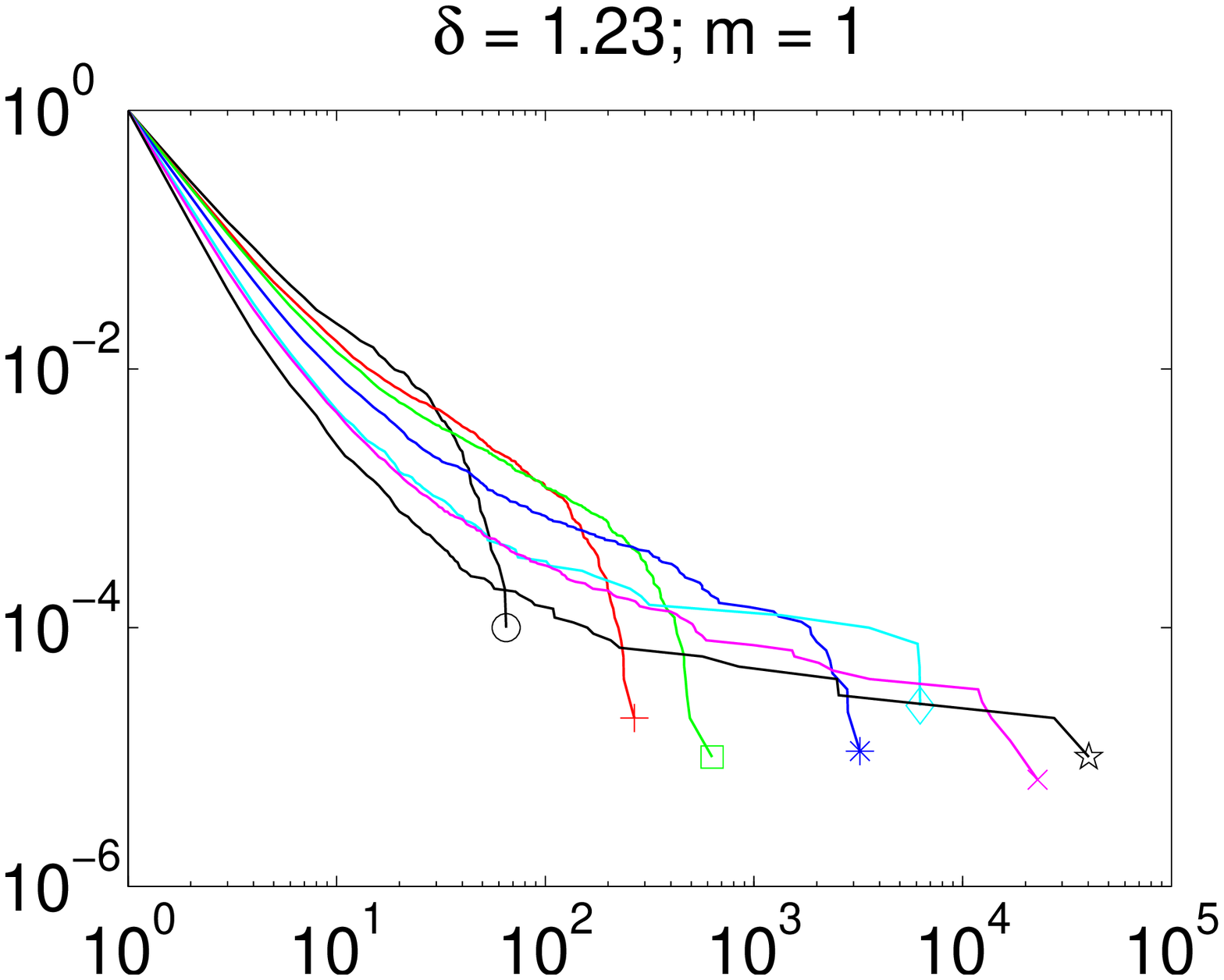}}\hfill
    \subfigure{\includegraphics[height=1.41in]{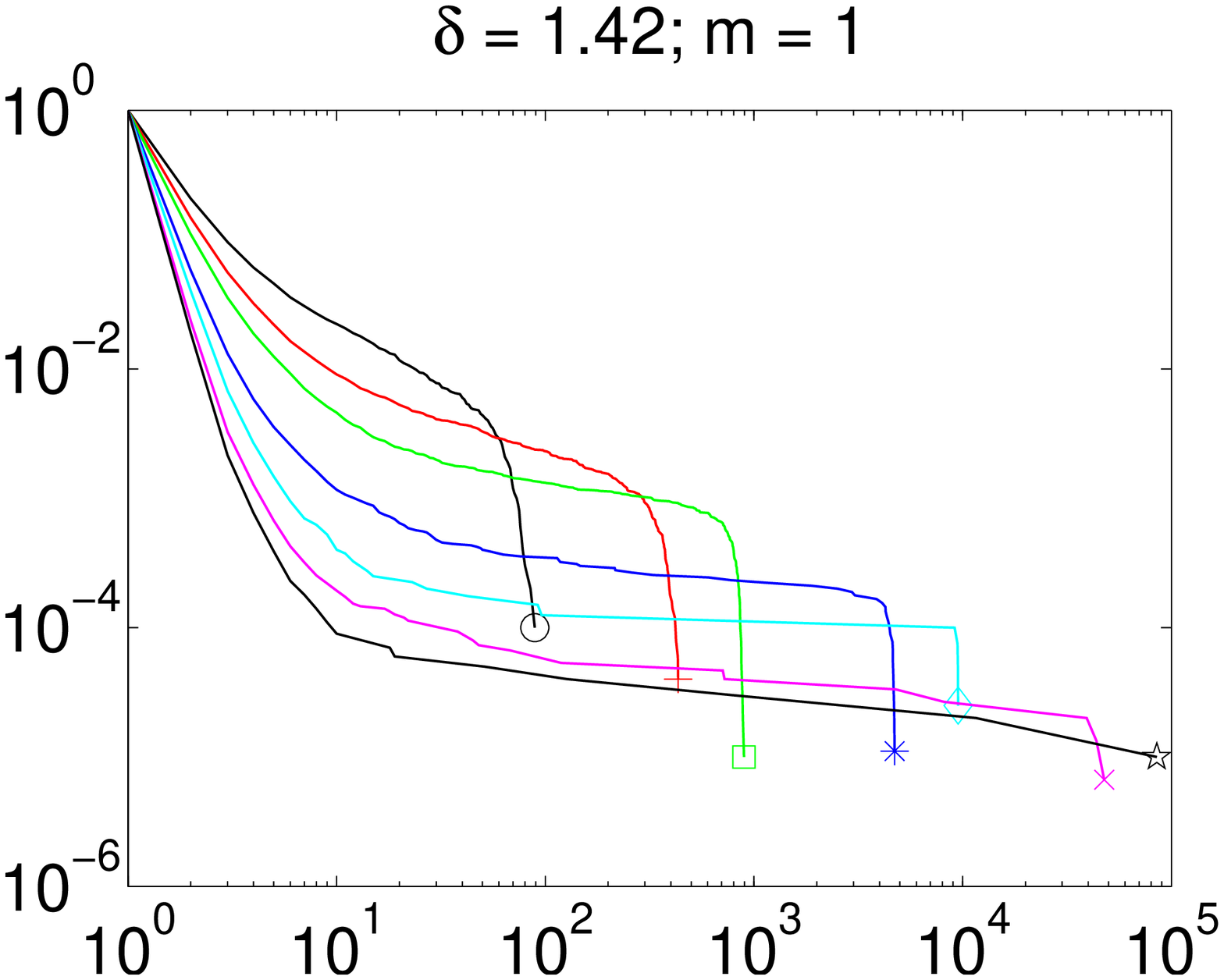}}\hfill
    \subfigure{\includegraphics[height=1.41in]{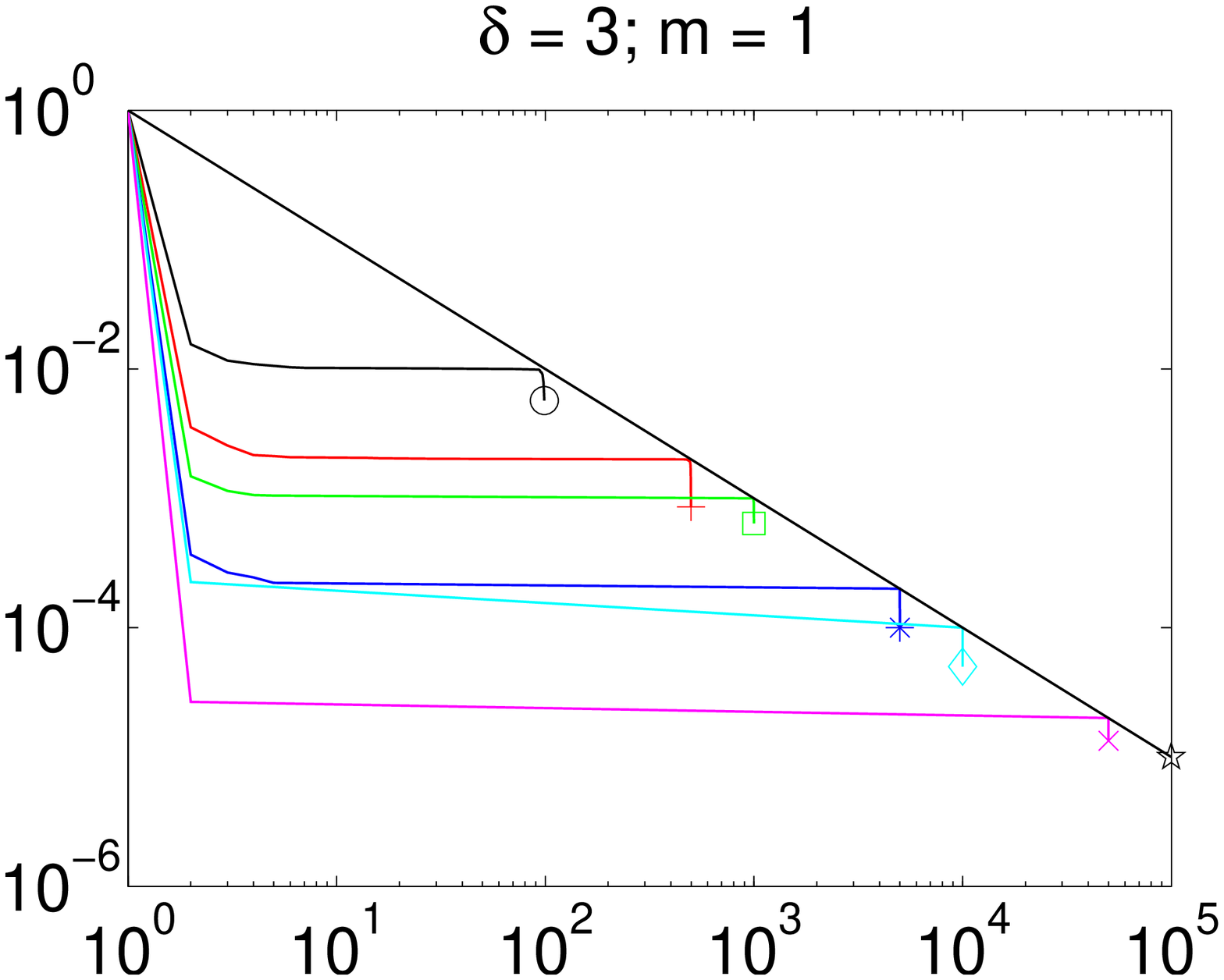}}\\
    \subfigure{\includegraphics[height=1.41in]{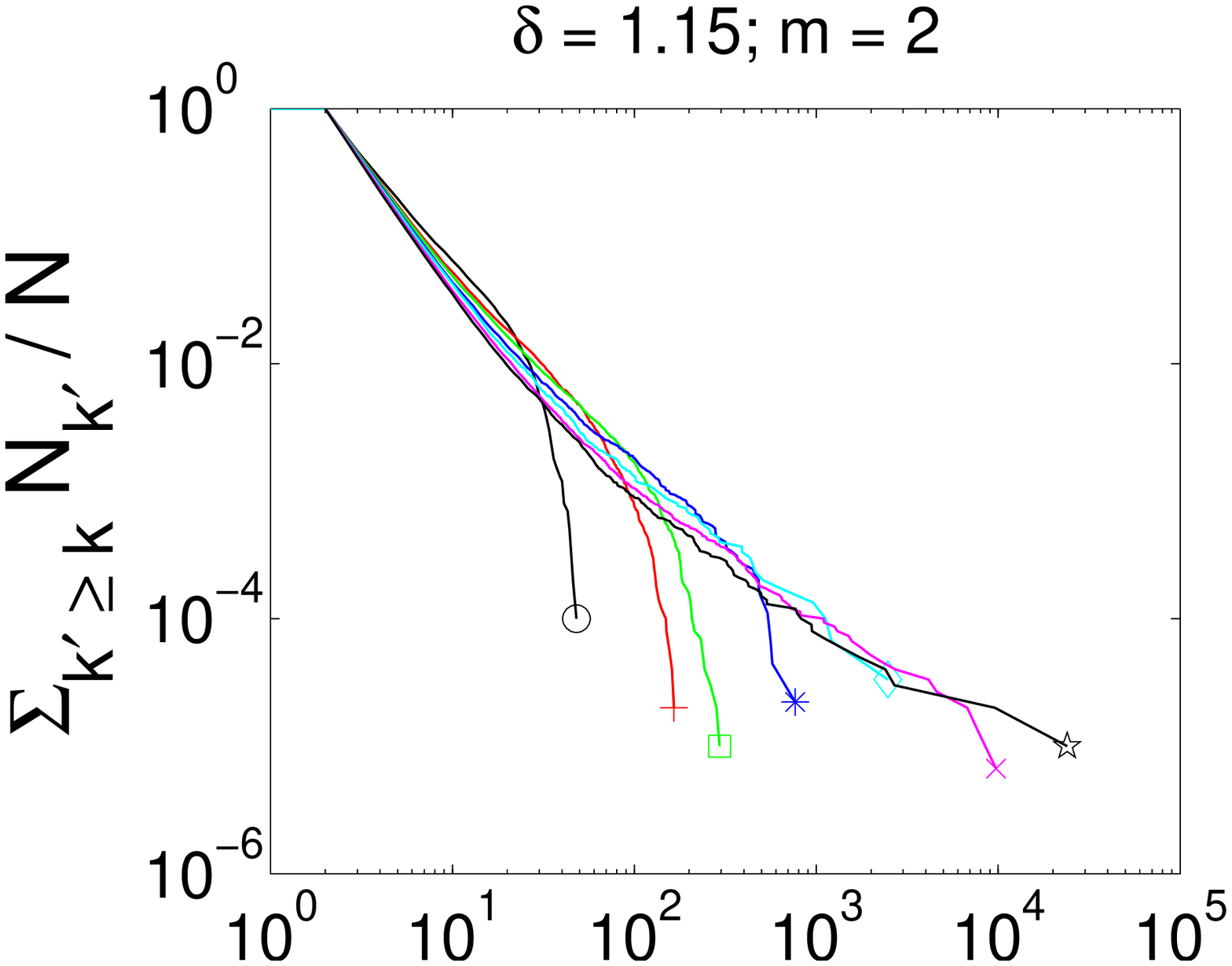}}\hfill
    \subfigure{\includegraphics[height=1.41in]{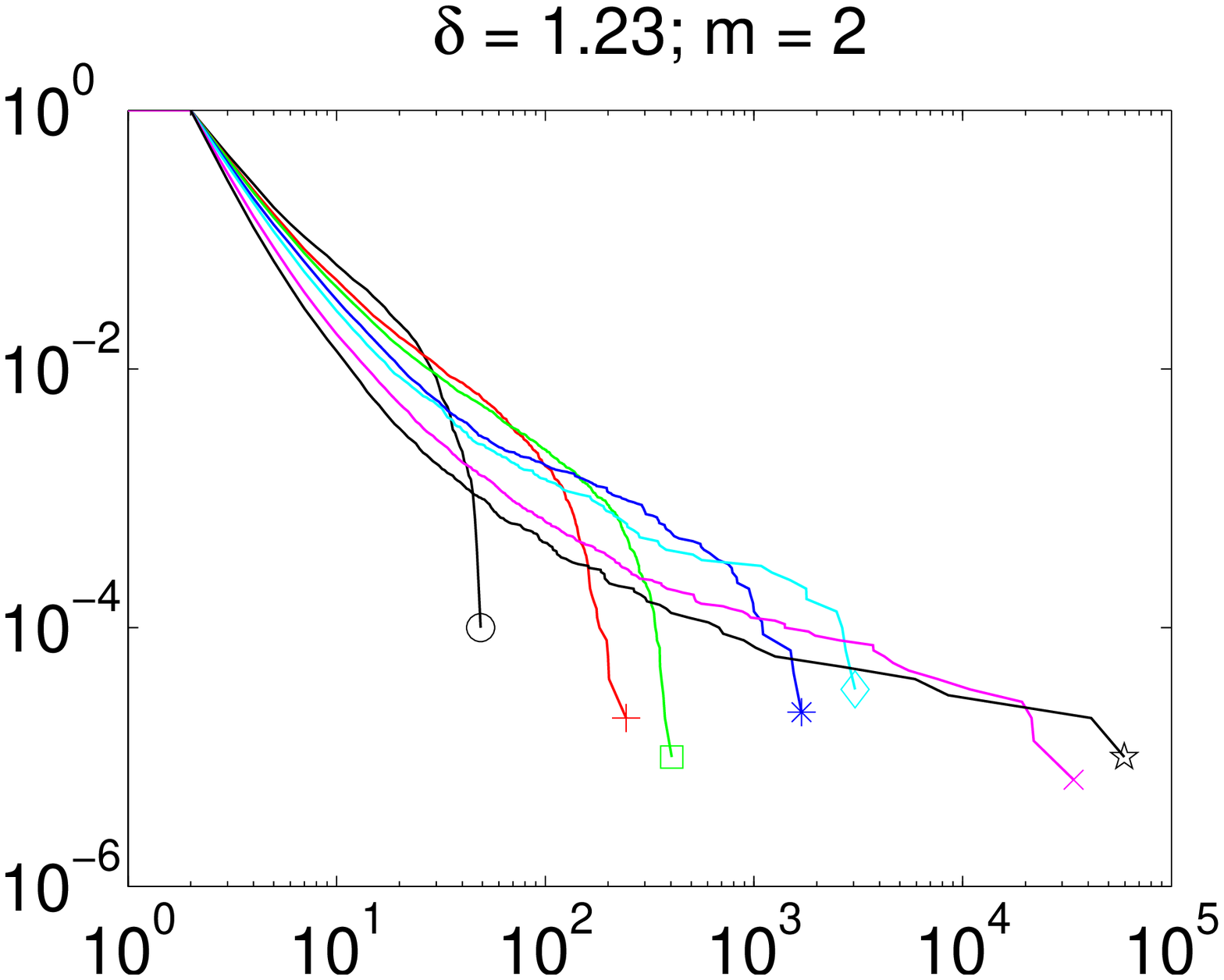}}\hfill
    \subfigure{\includegraphics[height=1.41in]{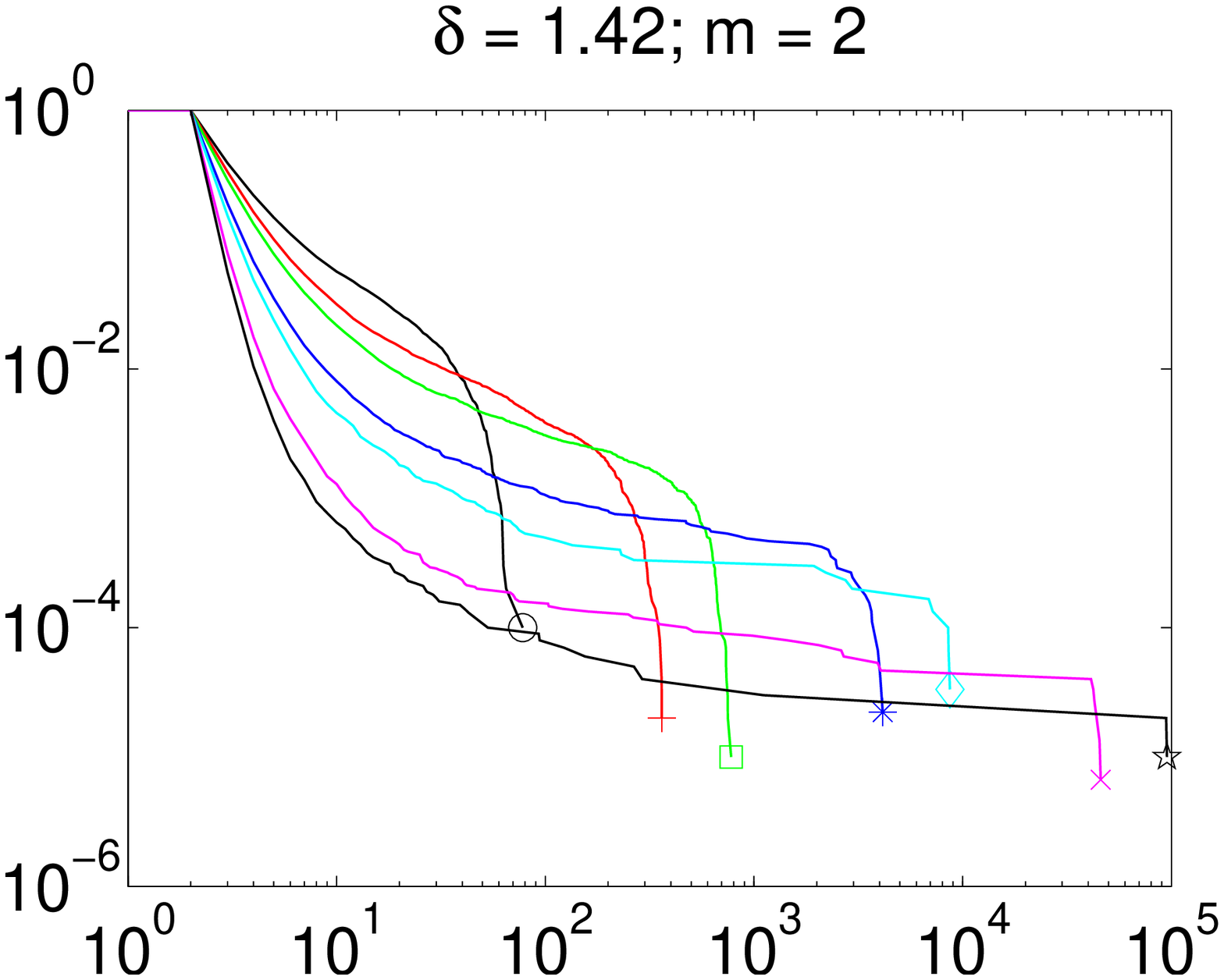}}\hfill
    \subfigure{\includegraphics[height=1.41in]{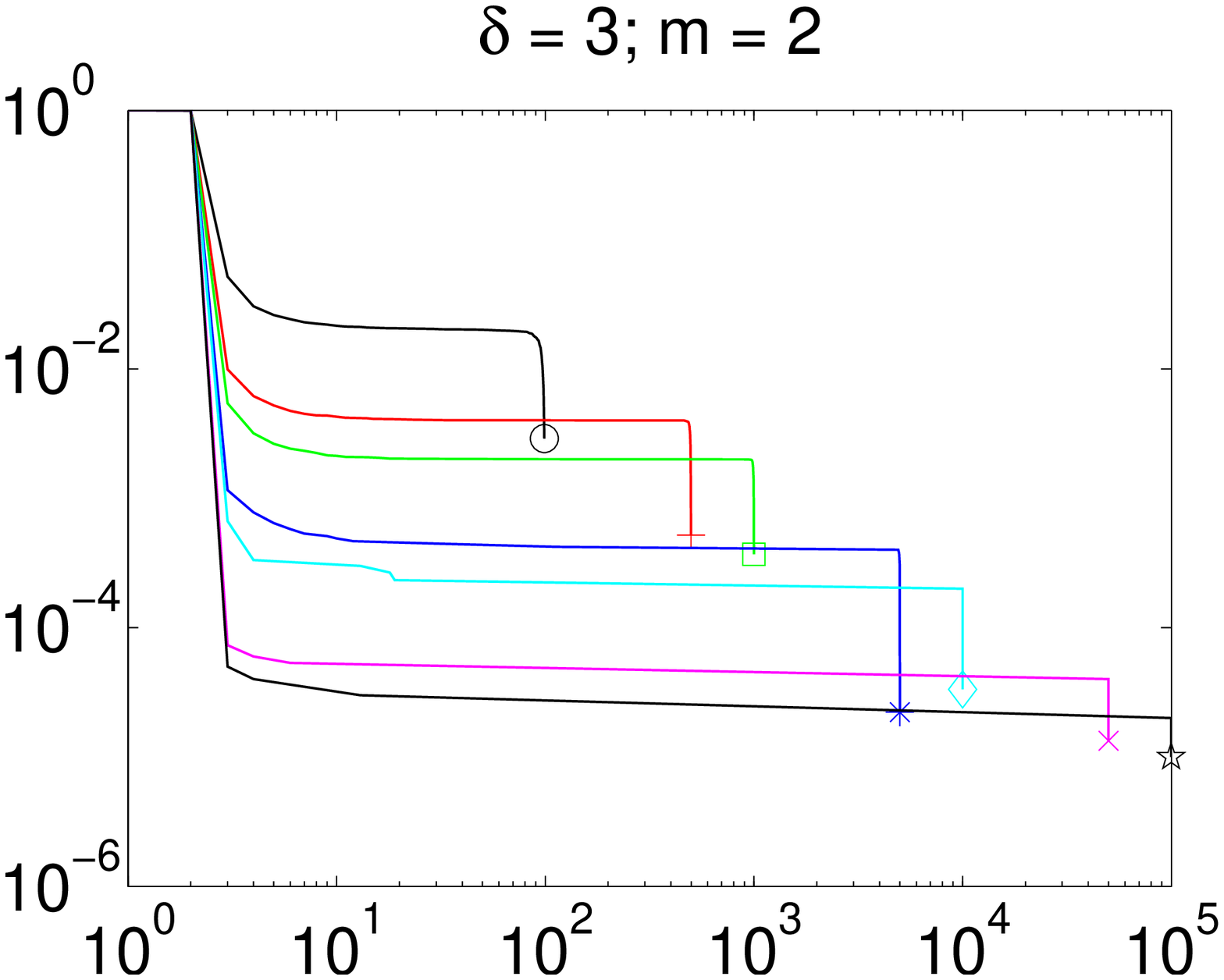}}\\
    \subfigure{\includegraphics[height=1.41in]{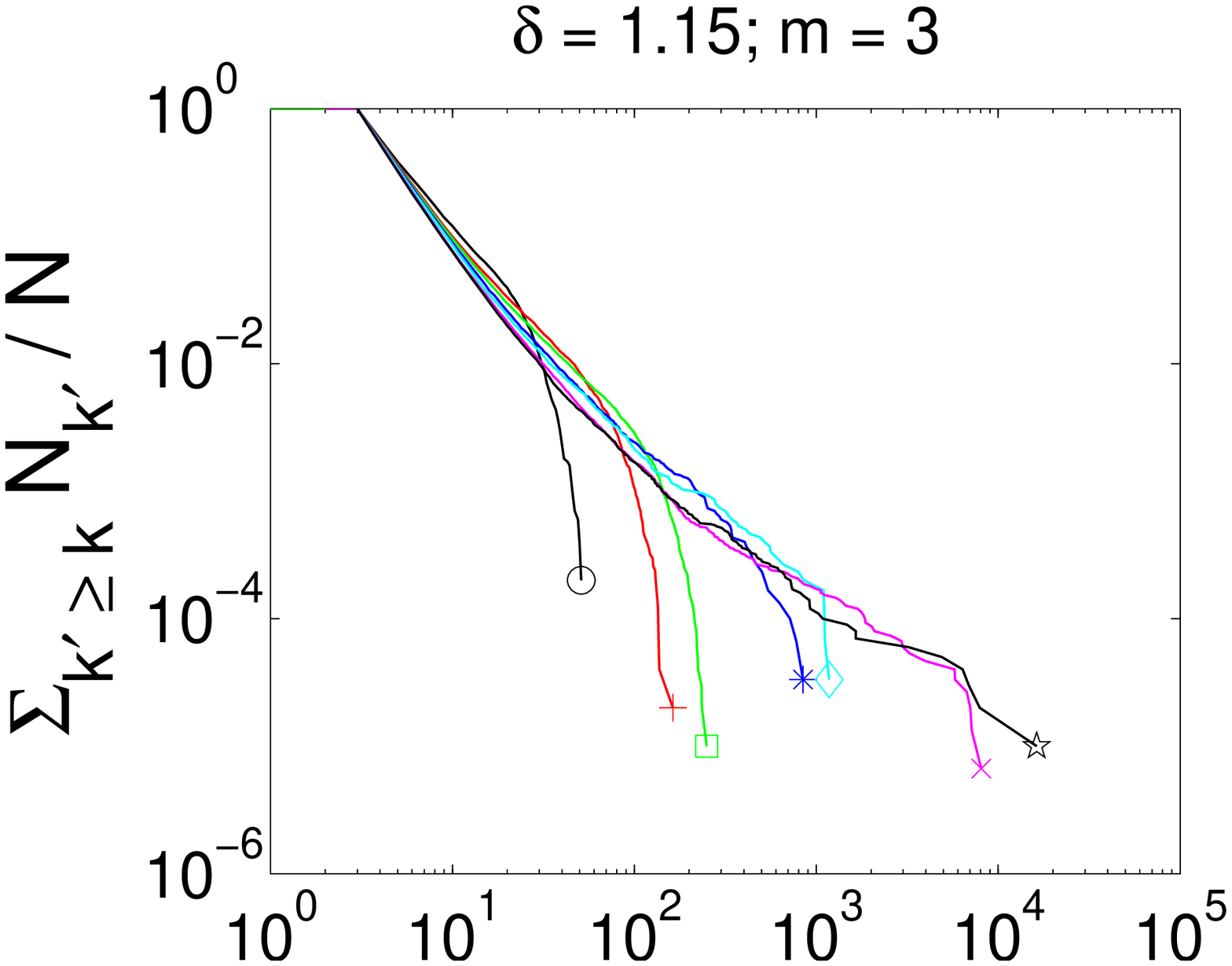}}\hfill
    \subfigure{\includegraphics[height=1.41in]{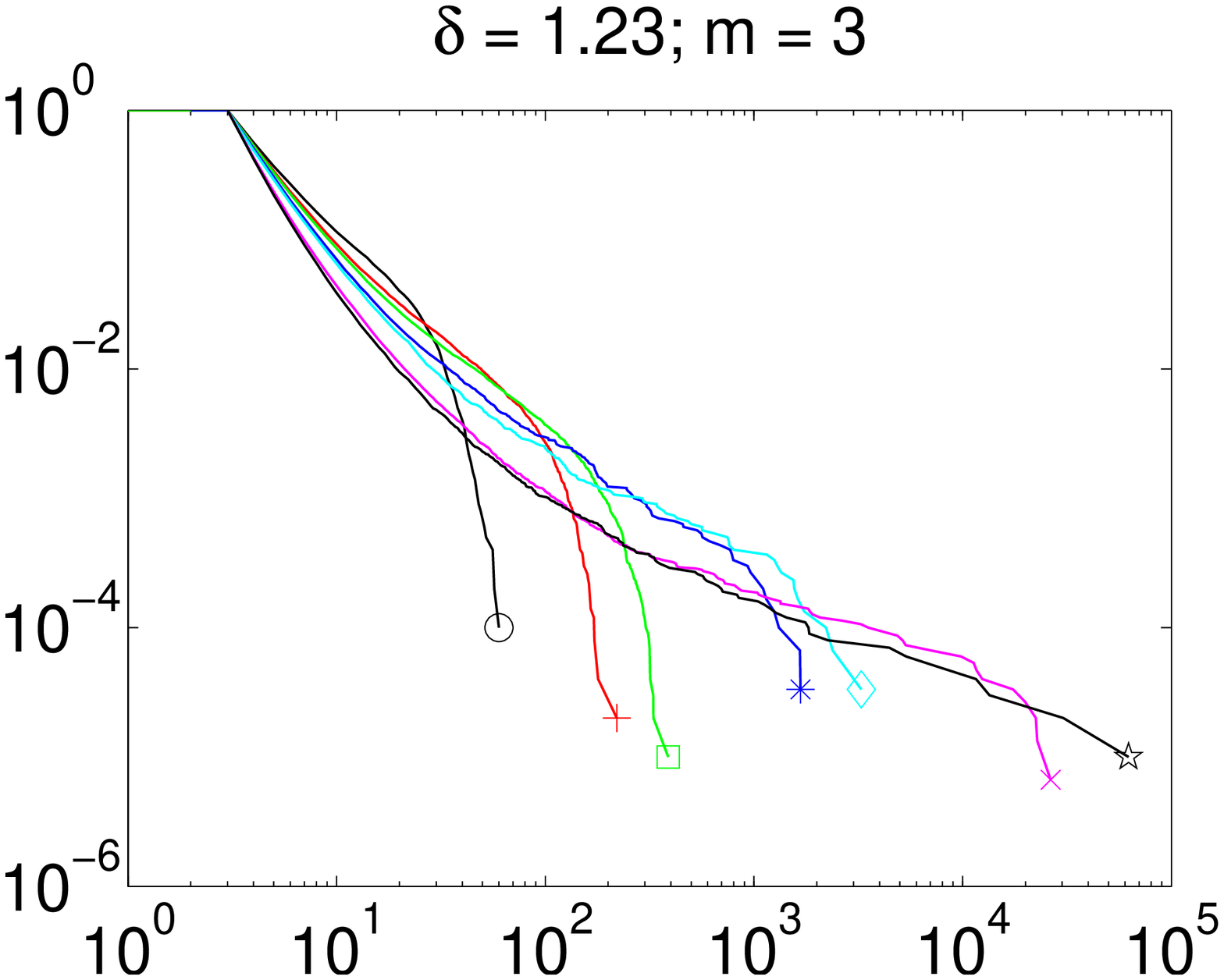}}\hfill
    \subfigure{\includegraphics[height=1.41in]{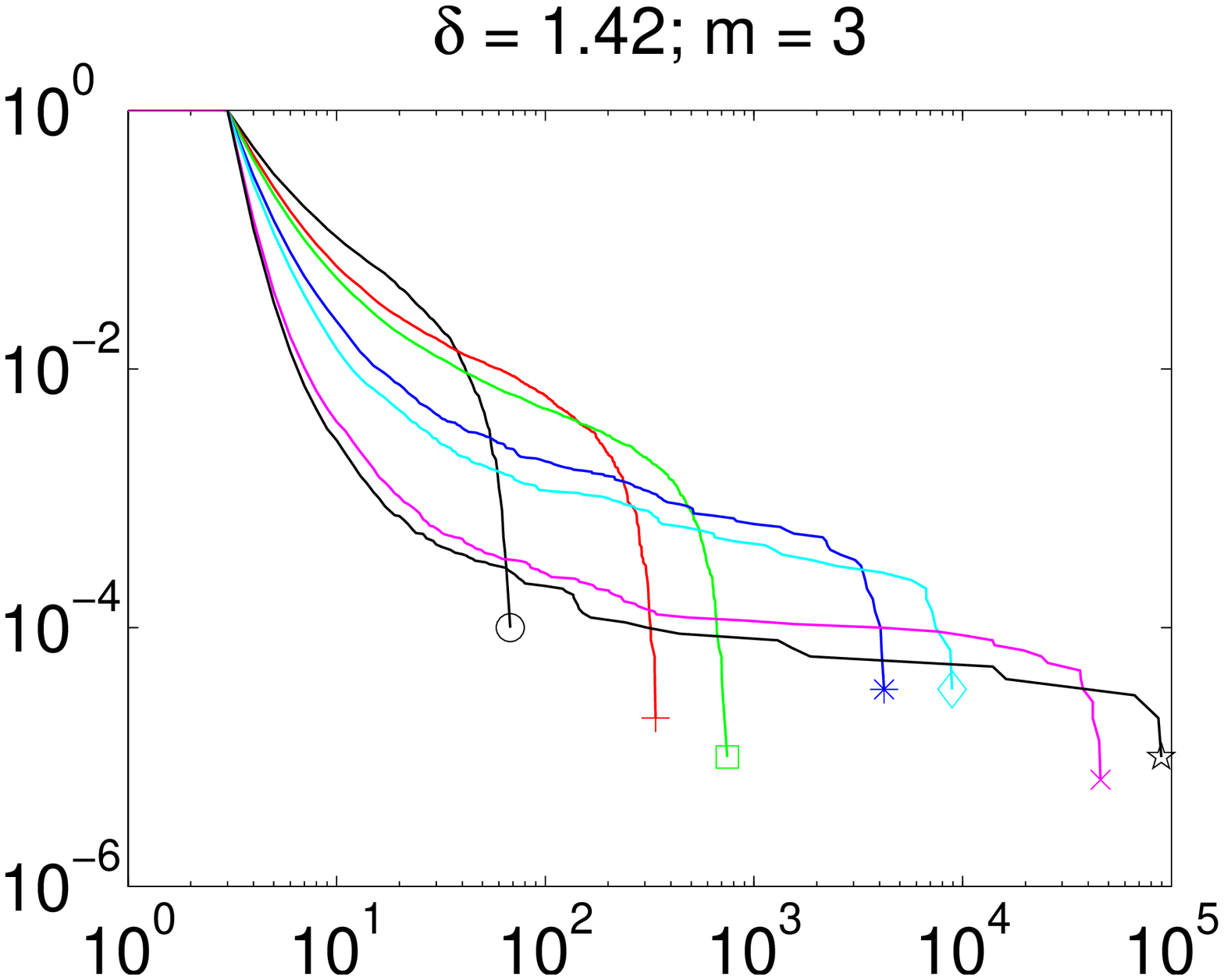}}\hfill
    \subfigure{\includegraphics[height=1.41in]{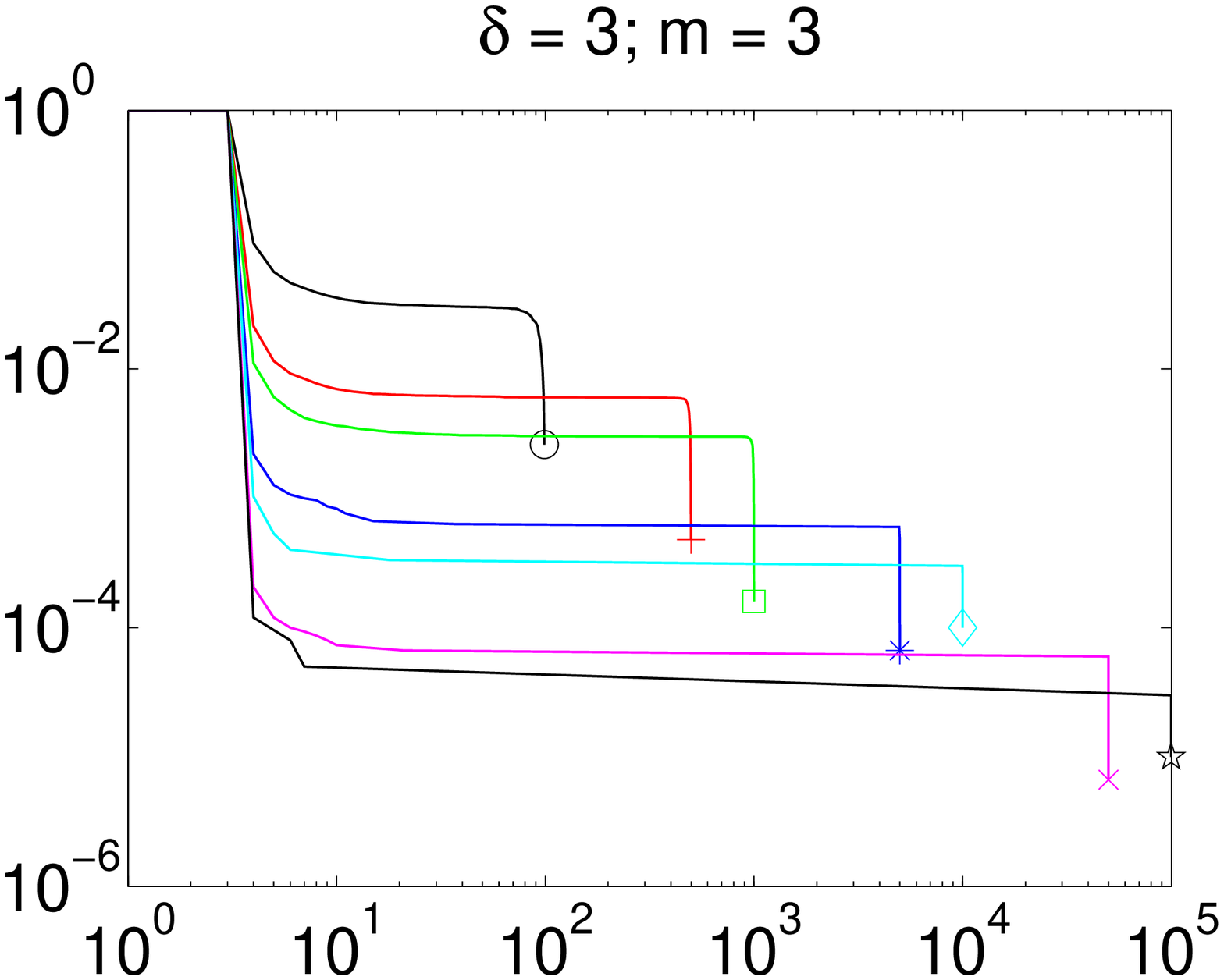}}\\
    \subfigure{\includegraphics[height=1.44in]{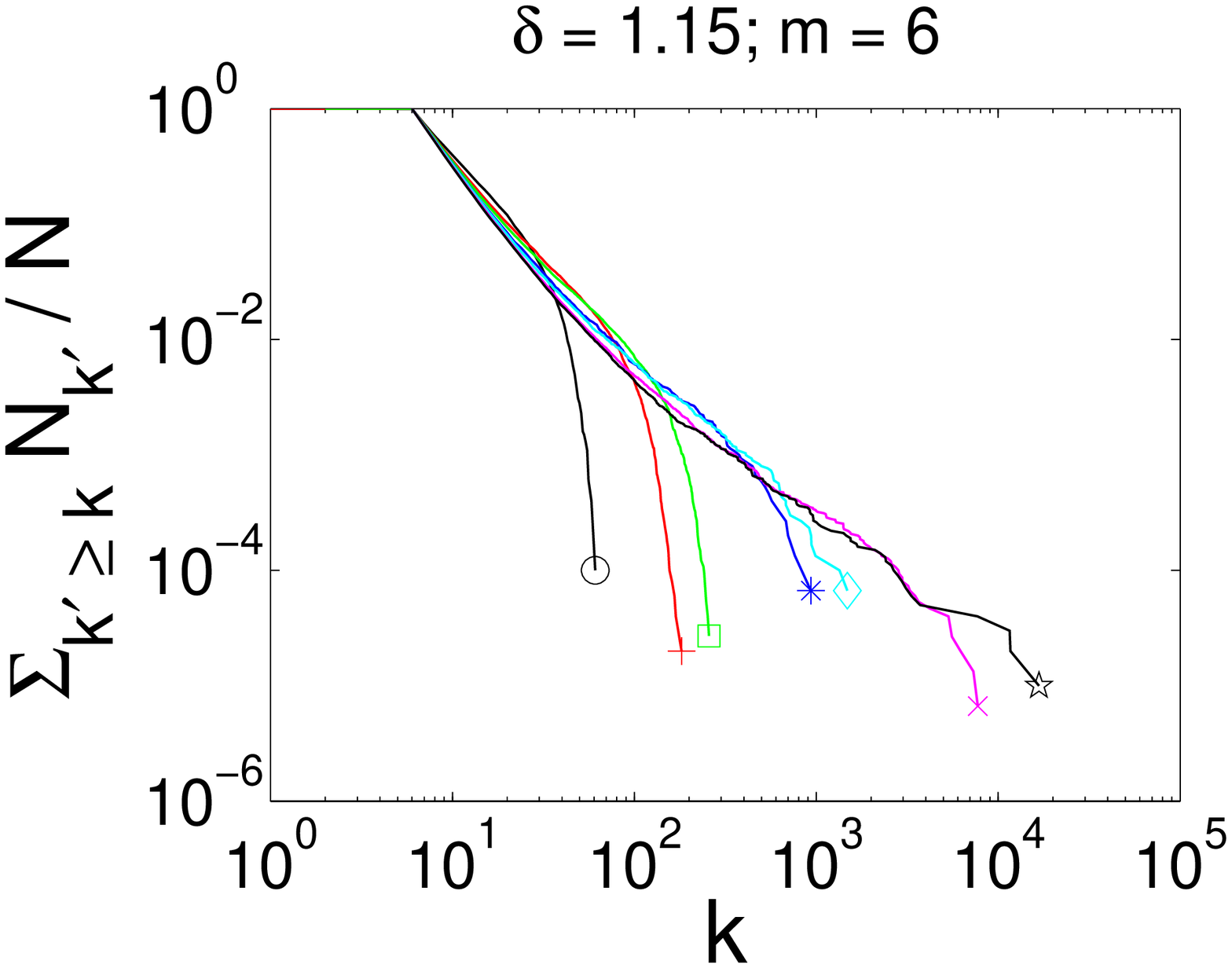}}\hfill
    \subfigure{\includegraphics[height=1.44in]{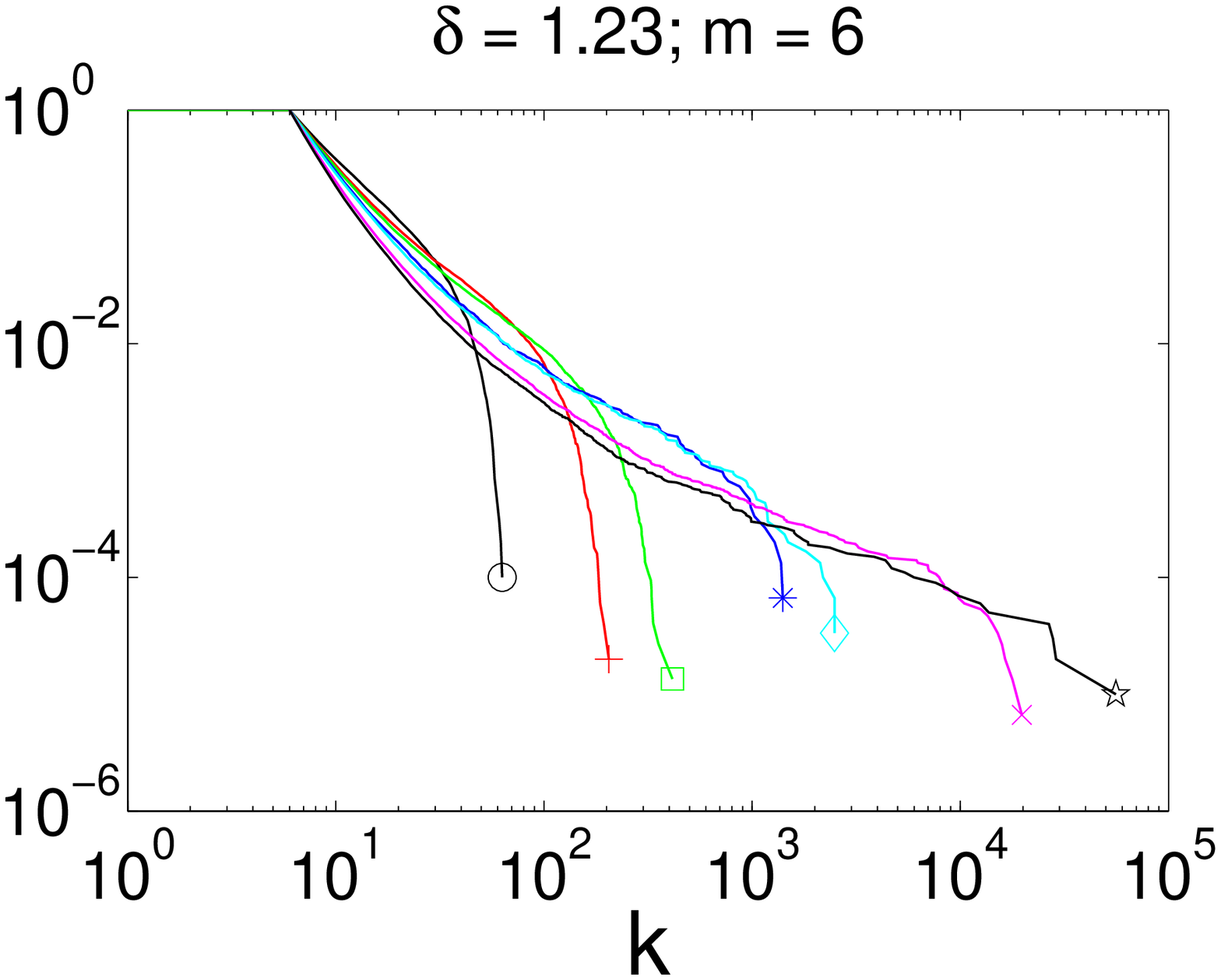}}\hfill
    \subfigure{\includegraphics[height=1.44in]{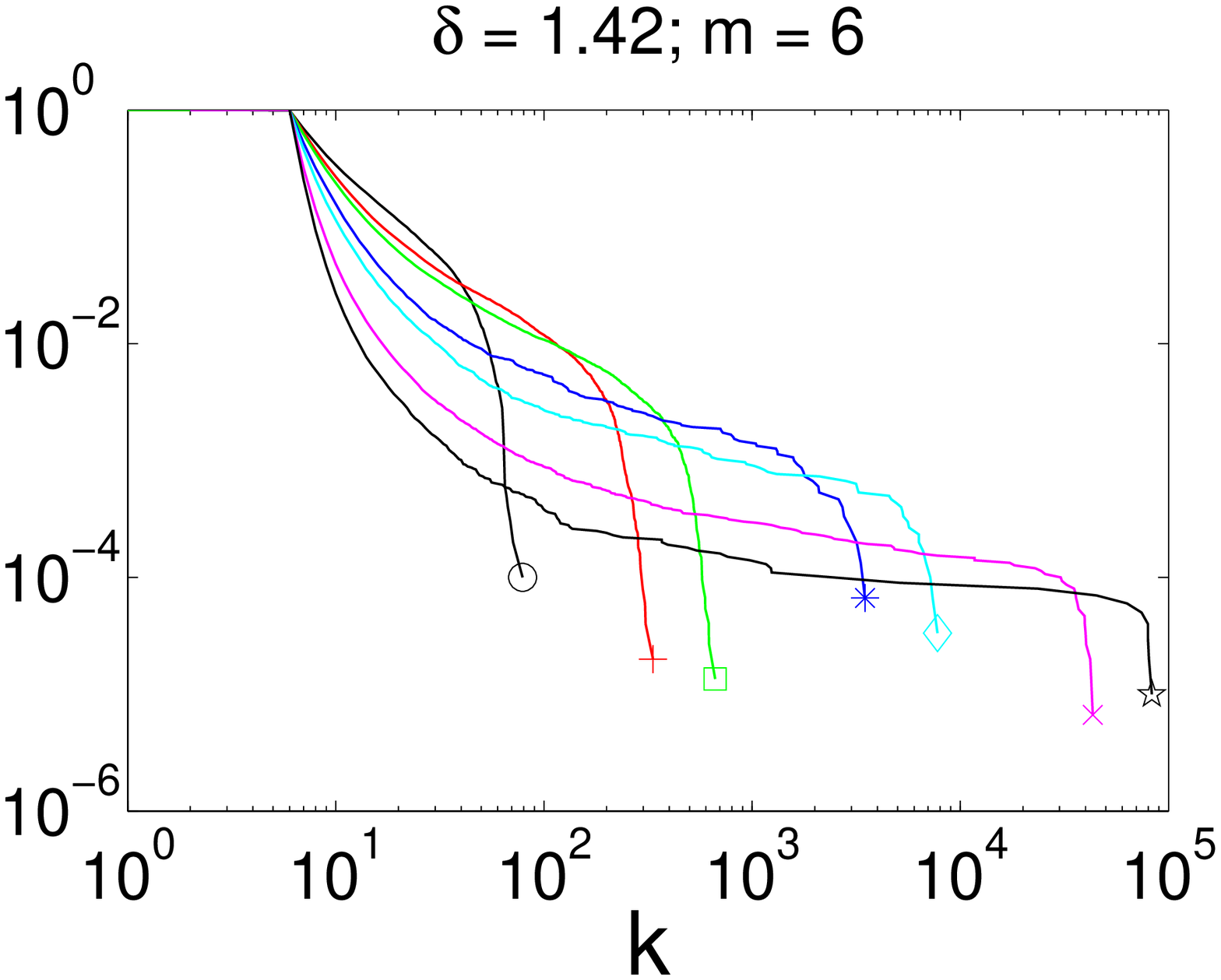}}\hfill
    \subfigure{\includegraphics[height=1.44in]{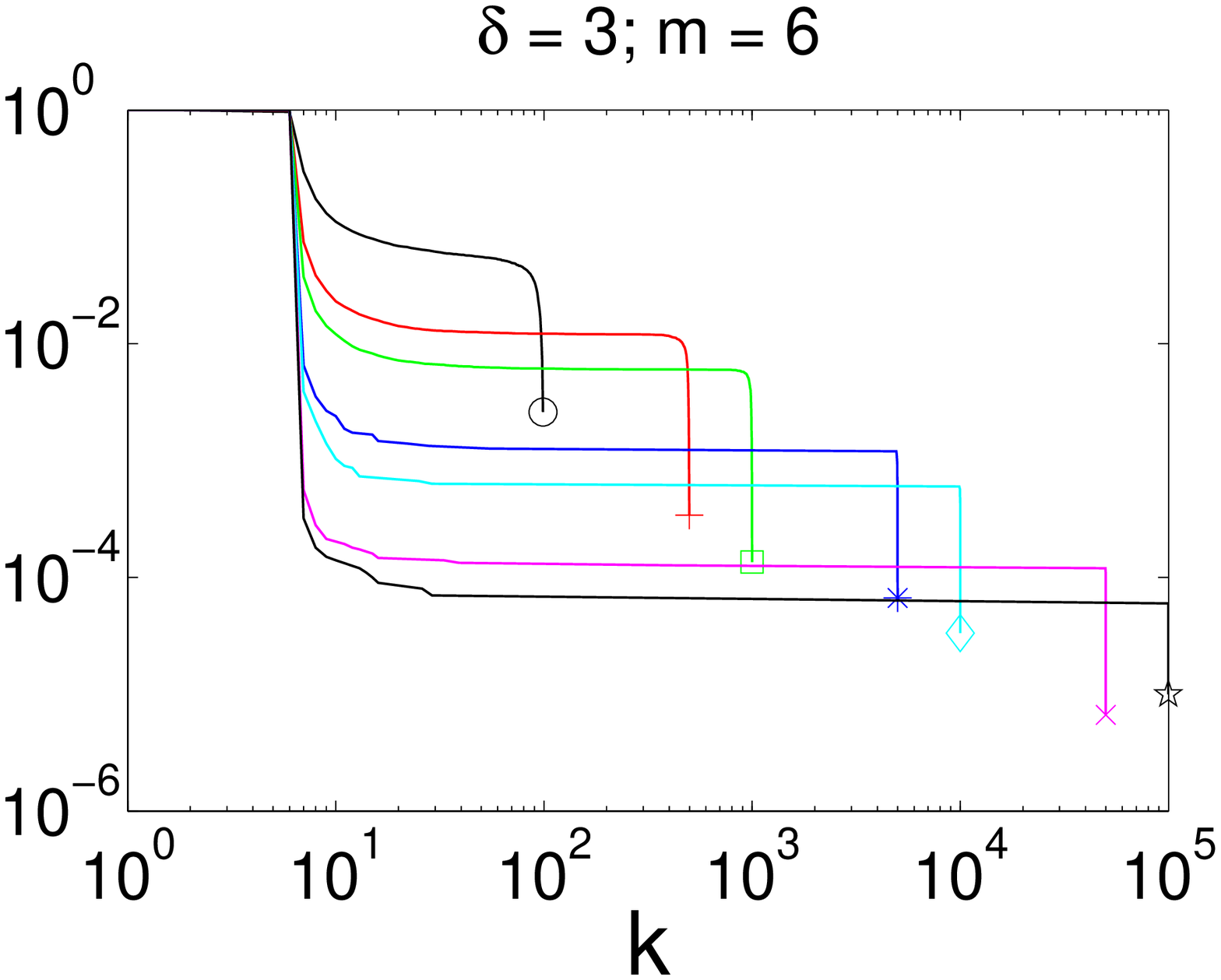}}
    \caption{(Color online) Scaling of the degree distributions
    in SLGNs with different $\delta$ and $m$. The lines show the
    complementary cumulative distribution of node degrees ($\sum_{k' \geq k}N_{k'}/N$)
    measured in simulations.}
    \label{fig:big}
\end{figure*}

We juxtapose these analytic estimates with simulations in
Figs.~\ref{fig:N3/N} and~\ref{fig:big}, showing the proportion of
degree-$3$ nodes $N_3/N$ and the overall degree distribution
$\sum_{k' \geq k}N_{k}/N$ in SLGNs of different size $N$, grown with
different $\delta$ and $m$. For each combination of $(N,\delta,m)$
we average the results over a number of graph instances ranging from
$3$ for the largest size $N=10^5$ to $100$ for smaller $N$. We
select the values of $\delta=(\delta_p+\delta_{p-1})/2$, $\delta_p =
1 + 1/p$, $p=1,\ldots,7$ ($\delta=3$ for $p=1$), so that the
selected $\delta$-values lie within the connectivity transition
intervals discussed above. For $p=7$, $\delta=1.15$, i.e., the
$\delta$-value used in \cite{ZhoMo04}.

In Fig.~\ref{fig:N3/N} the cases with $m=1$ and $m=2$ confirm the
expected: the larger $\delta$, the more quickly the proportion
$N_3/N$ approaches our analytic prediction of its asymptotic
scaling. Comparing $m=1$ and $m=2$, we see that in the former case,
only for $\delta=1.15$ does $N_3/N$ stay constant for all graph
sizes $N$ achieved in our simulations, while in the latter case
($m=2$), this ratio is constant for higher $\delta$-values ($\delta=1.23$) as
well. We see that for small $\delta$'s, the scalings of $N_3/N$ are
much farther from their asymptotes in the $m=2$ case than in the
$m=1$ case. The $m=3$ plot confirms that $N_3/N$ quickly saturates
to a dependent constant that increases with $\delta$, while for
$m=6$, $N_3/N$ decays as expected, $\sim1/N$, with the stronger
fluctuations, the smaller $\delta$.

Two factors contribute to the discrepancies between
the analytic predictions and simulations in Fig.~\ref{fig:N3/N}.
First, we neglected loss terms in Eq.~(\ref{N3-sol}).
Taking those into account would yield, for $m=2$, the asymptotic
expansion
\begin{equation}
\label{asymp-exp}
N_3(N)=a\,N^{2-\delta}-b\,N^{3-2\delta}+c\,N^{4-3\delta}+\ldots,
\end{equation}
where $b$ and $c$ are some constants that depend on $\delta$. For
$\delta=1.15$ this expansion turns into
\begin{equation}
N_3(N)=a\,N^{0.85}-b\,N^{0.7}+c\,N^{0.55}+\ldots
\end{equation}
explaining why keeping only the leading term in the asymptotic
result may lead to huge errors for small $\delta$ and $N$.

The second discrepancy factor is that
all the $N_k(N)$ analytic estimates above are actually the average
values of the corresponding random quantities. Nothing is known
about fluctuations of the degree distribution, the analysis of
which is difficult even in the simpler case of linear preferential
attachment~\cite{KraRe02}.

Fig.~\ref{fig:big} provides a more global view of the dependency of
the degree distribution on $\delta$ and $m$. The higher $\delta$,
the more skewed the degree distribution and hence the more star-like
the graphs. For $N=10^5$, $\delta=3$, and $m=1$, all the graph
instances in our simulations are stars. The larger $m$, the closer the degree
distribution curves corresponding to different $N$ are to each other
(neglecting the size-dependent cut-offs exhibited by all graphs), the
straighter these lines, and thus the weaker the dependency of the
degree distribution shape on the network size, and the deeper the
preasymptotic regime.

\section{Rich club connectivity versus joint degree distribution}
\label{seq:rcc-vs-jdd}

We have shown in the previous section that the power laws
empirically observed in the PFP model do not contradict the
asymptotic open book structure of SLGNs, since typical network sizes
considered in simulations are preasymptotically small. However,
this argument does not explain why the PFP model almost exactly
reproduces not only the power-law degree distribution observed in
the real Internet, but also a long list of other important network
properties. Since the preasymptotic regime is not amenable to
straightforward analytic treatment, in this section we approach the
problem from a different angle, and provide a simple explanation
based mostly on previous empirical work.

We first notice that the fact that the PFP model exhibits
preasymptotic power-law behavior is not so much surprising, because
for $\delta=1$ the model produces asymptotic power laws, and this
asymptote is quickly achieved for small $N$. The results of the
previous section indicate that if $\delta\gtrapprox1$ and $m>1$,
then this power-law asymptotic behavior unnoticeably changes to
preasymptotic, slowly transforming into the new asymptotic behavior
only for very large $N$.

Yet this argument does not explain why the PFP model reproduces so
many other network properties observed in the Internet. Previous
work~\cite{MaSneZa04,BiCaCa05,MaKrFo06,MaKrFaVa06-phys,SerKriBog07}
shows that the degree distribution alone does not fully define all
other Internet's properties, i.e., the Internet is not $1K$-random
in the terminology of \cite{MaKrFaVa06-phys}, but is almost
$2K$-random --- its structure is very close to the structure of
maximally random graphs constrained by its $2$-point degree
correlations, or the joint degree distribution (JDD) defined by the
total number $N_{kk'}$ of links between degree-$k$ and degree-$k'$
nodes.
In other words, the Internet's JDD narrowly defines almost all its
other important properties, except clustering
\cite{MaKrFaVa06-phys,SerKriBog07}.

Although the PFP model is not
concerned with the JDD {\it per se}, it
reproduces precisely the observed rich club connectivity (RCC)
$\varphi(r/N)$ defined as the ratio of the number of links in the
subgraph induced by the $r$ highest degree nodes to the maximal
number of such links ${r \choose 2}$. The values of $\varphi(r/N)$
observed in the Internet for small $r$ are substantially higher than
in networks grown according to linear preferential attachment.
Superlinear preference increases the connectivity density among high-degree
nodes, which explains why the PFP model successfully captures the observed RCC.

In the rest of this section we
analyze the relationship between the JDD and RCC. Specifically, the
JDD almost fully defines RCC: any two graphs with the same JDD have
almost the same RCC. While the converse is generally not true, a
given form of RCC introduces certain constraints to the JDD. Given
the JDD's definitive role for the Internet topology, we conclude
that reproducing Internet's RCC must significantly improve the accuracy
in capturing all other properties of the Internet topology that
depend on degree correlations, which explains the success of the PFP
model and provides clear grounds for the discussion
in~\cite{Zhou06,ZhoMo07}.

To see that the JDD almost fully defines RCC is
straightforward~\cite{CoFlSeVe06}. We first get rid of the node rank
$r$ in $\varphi(r/N)$. The rank of a node is its position in the
degree sequence sorted in decreasing order, i.e., as in
(\ref{book}). Recall that the node rank is essentially the
complementary cumulative distribution function for node degrees: if
$d_i$ and $r_i$ are the degree and rank of node $i$, $k_{\max}$ is
the maximum degree, and if we denote $N_k^+ = \sum_{k'=k}^{k_{\max}}
N_{k'}$, then $1+N_{d_i+1}^+ \leq r_i \leq N_{d_i}^+$. Thus, the JDD
and RCC are directly related via $\varphi_k$ defined as the total
number of links between degree-$k$ nodes and nodes $i$ of higher
degrees $d_i \geq k$
\begin{eqnarray}
\label{eq:rcc_vs_jdd}
\varphi_k &=& {N_k^+ \choose 2} \varphi(N_k^+/N) -
{N_{k+1}^+ \choose 2} \varphi(N_{k+1}^+/N)\nonumber\\
&=& \sum_{k'=k}^{k_{\max}} N_{kk'}.
\end{eqnarray}
It follows that the JDD defines RCC, up to reordering of nodes of
the same degree.

To illustrate how the RCC constrains JDD, we choose to consider a
common projection of the JDD, the average degree of the nearest
neighbors of degree-$k$ nodes $\bar{k}_{nn}(k)$. We first look at
the maximum and minimum possible value of $\bar{k}_{nn}(k)$ for a
class of graphs with some fixed degree distribution with minimum
and maximum degrees of $1$ and $k_{\max}$. We then suppose that
$\varphi_k$ is also given as a constraint, and we quantify how this
constraint narrows down the spectrum of possible values of
$\bar{k}_{nn}(k)$.

It is easy to see that the minimum and maximum values of
$\bar{k}_{nn}(k)$ without the $\varphi_k$ constraints are simply $1$
and $k_{\max}$, if we neglect any structural constraints that a
given form of the degree distribution imposes on possible JDDs. For
example, if $N_1>k_{\max}N_{k_{\max}}$, then $\bar{k}_{nn}(1)$
cannot be $k_{\max}$, it is necessarily less than $k_{\max}$.
Scale-free networks with $\gamma<3$ have these constraints for links
connecting nodes of degrees $k$ and $k'$ such that $kk'>\bar{k}N$
\cite{BoPaVe04}. To formally see that without such constraints the
minimum and maximum of $\bar{k}_{nn}(k)$ is $1$ and $k_{\max}$, let
$\mu_{kk'}=1+\delta_{kk'}$ be the factor taking care of links
between nodes of the same degree in $M_{kk'}=\mu_{kk'}N_{kk'}$, so
that the total number $M_k$ of ``edge ends'' (stubs) attached to
degree-$k$ nodes is $M_k=kN_k=\sum_{k'}M_{kk'}$. We then have, by
definition,
\begin{equation}
\label{eq:knn_def}
 \bar{k}_{nn}(k) = \frac{1}{M_k}\sum_{k'}k'M_{kk'}.
\end{equation}
(The more common definition for the normalized distributions
$P(k)=N_k/N$ and $P(k,k')=M_{kk'}/(\bar{k}N)$ such that
$\sum_kP(k)=\sum_{kk'}P(k,k')=1$ is
$\bar{k}_{nn}(k)=\sum_{k'}k'P(k'|k)=\bar{k}/(kP(k))\sum_{k'}k'P(k',k)$.)
The minimum (maximum) values of~$\bar{k}_{nn}(k)$ are achieved when
all degree-$k$ nodes are attached only to the nodes with the minimum
(maximum) degrees,
\begin{eqnarray*}
\bar{k}_{nn}^{\min}(k) &=& \frac{1}{M_k}
\min\left(\sum_{k'}k'M_{kk'}\Big|\sum_{k'}M_{kk'}=M_k\right)= 1,\\
\bar{k}_{nn}^{\max}(k) &=& \frac{1}{M_k}
\max\left(\sum_{k'}k'M_{kk'}\Big|\sum_{k'}M_{kk'}=M_k\right)= k_{\max},\\
\end{eqnarray*}
where the minimum (maximum) is taken over all possible JDD matrices
$M_{kk'}$ yielding the given degree distribution $M_k$. We thus see
that the maximum difference between possible values
of~$\bar{k}_{nn}(k)$ is
\begin{equation}
\label{eq:delta_nophi} \Delta(k) = \bar{k}_{nn}^{\max}(k) -
\bar{k}_{nn}^{\min}(k) = k_{\max}-1.
\end{equation}
In the Internet the maximum node degree is large (it scales as
$k_{\max} \sim N^{\frac{1}{\gamma-1}}$ \cite{BoPaVe04}), and hence
$\Delta(k) \approx k_{\max}$.

Suppose now that $\phi_k=\sum_{k'=k}^{k_{\max}}M_{kk'}=\varphi_k+N_{kk}$
is given as a constraint. Note that $\phi_k$ is not precisely
equal to $\varphi_k$, but we neglect this extra $N_{kk}$ term here
as well, partly because in the Internet, $N_{kk}$ is relatively
small for almost all $k$. Introducing ratio $\alpha_k=\phi_k/M_k$,
which is approximately the ratio of the number of links connecting
degree-$k$ nodes and nodes of higher degrees to the number of all
links attached to degree-$k$ nodes, we write the new minimum value
of $\bar{k}_{nn}(k)$ as
\begin{widetext}
\begin{eqnarray}
\bar{k}_{nn}^{\min}(k|\alpha_k)
    &=& \frac{1}{M_k} \left\{
    \min\left(
    \sum_{k'=1}^{k-1}k'M_{kk'}\Big|\sum_{k'=1}^{k-1}M_{kk'}=M_k-\phi_k
    \right)
    +
    \min\left(
    \sum_{k'=k}^{k_{\max}}k'M_{kk'}\Big|\sum_{k'=k}^{k_{\max}}M_{kk'}=\phi_k
    \right) \right\}\nonumber\\
    &=& \frac{1}{M_k} \left\{ 1 \cdot (M_k-\phi_k)
        + k \cdot \phi_k \right\}
    = (k-1)\alpha_k + 1,
\end{eqnarray}
\end{widetext}
where the minimum is now taken over all JDDs $M_{kk'}$ satisfying
the RCC constraints. Similarly, for the maximum possible value, we have
\begin{eqnarray}
\bar{k}_{nn}^{\max}(k|\alpha_k)
    &=& \frac{1}{M_k} \left\{ (k-1) \cdot (M_k-\phi_k)
        + k_{\max} \cdot \phi_k \right\}\nonumber\\
    &=& (k_{\max}-k+1)\alpha_k + k-1,
\end{eqnarray}
and the maximum possible difference is
\begin{eqnarray}
\label{eq:delta_phi} \Delta(k|\alpha_k) &=&
\bar{k}_{nn}^{\max}(k|\alpha_k)
- \bar{k}_{nn}^{\min}(k|\alpha_k)\nonumber\\
&=& (k_{\max}-2k+2)\alpha_k + k-2.
\end{eqnarray}

\begin{figure}[tb]
     \includegraphics[width=.8\linewidth]{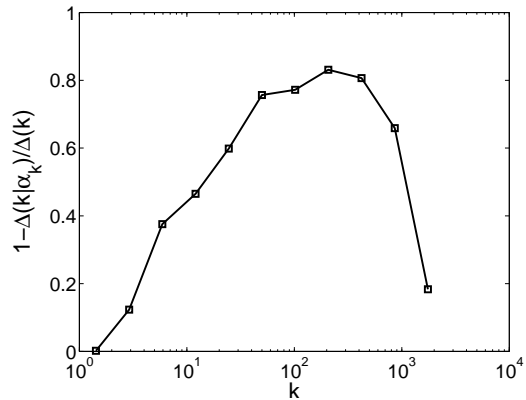}
  \caption{Relative decrease of the range of possible values of
  $\bar{k}_{nn}(k)$ imposed by the Internet's RCC. The Internet map
  from \cite{MaKrFo06} is used to compute $\Delta(k)$
  and $\Delta(k|\alpha_k)$ given by
  Eqs.~(\ref{eq:delta_nophi},\ref{eq:delta_phi}).}
\label{fig:deltak}
\end{figure}

Compared to the unconstrained case, the relative decrease of the
range of possible values of $\bar{k}_{nn}(k)$, assuming
large~$k_{\max}$, is
\begin{eqnarray}
\label{eq:delta_delta}
\frac{\Delta(k)-\Delta(k|\alpha_k)}{\Delta(k)}
&\approx& \left(1-\frac{k}{k_{\max}}\right)-\left(1-2\frac{k}{k_{\max}}\right)\alpha_k\nonumber\\
&\approx&
\begin{cases}
1-\frac{k}{k_{\max}} & \text{if $\alpha_k \approx 0$};\\
\frac{1}{2}   & \text{if $\alpha_k \approx \frac{1}{2}$};\\
\frac{k}{k_{\max}}   & \text{if $\alpha_k \approx 1$}.
\end{cases}
\end{eqnarray}
In disassortative networks, such as the Internet \cite{MaKrFo06},
most links incident to medium- and high-degree nodes lead to
low-degree nodes, meaning that $\alpha_k \approx 0$ except for
$k/k_{\max} \ll 1$. Given (\ref{eq:delta_delta}), we conclude that
the RCC introduces significant constraints to the JDD, reflected even in a JDD's simple summary
statistic $\bar{k}_{nn}(k)$, except for lowest degrees $k \approx
0$, and perhaps highest degrees $k \approx k_{\max}$,
for which our analysis may be not very accurate since we neglected
the structural constraints that are relevant in the high-degree
zone. We confirm this conclusion in Fig.~\ref{fig:deltak} where we
use the RCC in the measured Internet topology to compute the
RCC-induced relative decrease $1-\Delta(k|\alpha_k)/\Delta(k)$ of
the range of possible values of $\bar{k}_{nn}(k)$. In the
medium-degree zone this decrease reaches 80\%.

\section{Conclusion}
\label{sec:conclusion}

Preferential attachment is a robust mechanism that may be
responsible for the emergence of the power-law degree distributions
in some complex networks~\cite{BarAlb99}. However,
power laws emerge only if the preference kernel is a linear
function of node degree~\cite{KraReLe00,KraRe01}. If one believes that
preferential attachment is a driving force, explicit or
implicit, behind the evolution of complex networks, then the natural
question one has to face is why this kernel must be exactly linear
in so many so different complex systems.

In this paper we argue that even if the preference kernel is not
linear but slightly superlinear, preferential attachment may still
produce scale-free networks, except that it does so not in the
asymptotic but in a vast preasymptotic regime. Two
key factors contribute to the depth of this regime: 1)~how close the
preference kernel is to being linear, and 2)~how many links are
added per new node. These factors allow us to say,
informally, that multiple links added under slightly superlinear
preferential attachment resurrect power laws, although only by means
of deepening the preasymptotic regime.

The asymptotic regime is still degenerate: adding $m$ links leads to the
asymptotic degree distribution $P(k)\to\delta_{k,m}$. More
precisely, the asymptotic network structure is a distorted (or
``torn'') open $m$-book --- a generalization of the known object in
topology~\cite{Giroux05}. The level of distortion depends on how
close the preference kernel is to a linear function. Similar to the
$m=1$ case (the open $1$-book is a star), we find an infinite series
of connectivity transitions characterizing the degree of damage to
the open book structure, as the kernel approaches a linear function.

To explain the success of one particular superlinear model --- the
positive-feedback preference model~\cite{ZhoMo04} --- in capturing
not only the degree distribution but also many other important
properties of one particular complex network, the Internet, we
analyze the both-way relationship between the joint degree
distribution ($2$-point degree correlations) and rich-club
connectivity. The former defines the latter, while the latter
constrains the former. These constraints, captured by the model,
suffice to reproduce many other important Internet's
properties, since it has been shown that most of them, except
clustering, depend only on the joint degree
distribution~\cite{MaKrFaVa06-phys,SerKriBog07}.

Given that the depth of the preasymptotic regime increases with the
number $m$ of links added per node, and that the average degrees
$\bar{k} \approx 2m$ of some complex networks including the Internet
have been reported to grow with network size
\cite{DoMe01,DoMe02,LeCleFa07}, our findings, taken altogether,
imply that some complex networks may exist in vast preasymptotic
regimes of evolution processes that have degenerate network
formations as their asymptotes. We contrast this implication with
the observation that the vast majority of the existing network
evolution models are designed with the goal to yield asymptotic
power-law distributions, quickly achievable at small network sizes.

An interesting open question is whether the dynamics of the world
economy supports our findings. Specifically, does the superlinear
growth of wealth contribute to such effects as the ``shrinking
middle class'' \cite{Winnick89-book,Ornstein07-book} and growing
wealth inequality \cite{DaSa07,DaSa08}? More succinctly, is the
Pareto distribution preasymptotic~\cite{DoMe02,BuJo02}?\\

\begin{acknowledgments}

We thank Mari{\'a}n Bogu{\~n}{\'a}, M.~\'Angeles Serrano, and kc
claffy for very useful discussions and suggestions. This work was
supported in part by NSF CNS-0434996 and CNS-0722070, by DHS
N66001-08-C-2029, and by Cisco Systems.

\end{acknowledgments}

\appendix*

\section{Nonextremal growth}
\label{sec:appendix}

The network remains an open book throughout its evolution with
probability
\begin{equation}
\mathcal{P}_\infty = \prod_{j=2}^\infty  \mathcal{P}_{j\mapsto j+1}
\end{equation}
where $\mathcal{P}_{j\mapsto j+1}$ is the probability of attaching
the new node to the two nodes of highest degree
\begin{equation*}
\mathcal{P}_{j\mapsto j+1} = \frac{(j-1)^\delta \cdot (j-2)^\delta}
{(j-1)^\delta + (j-2)^\delta + (j-3)\cdot 2^\delta +1}\,Q_j
\end{equation*}
and we used the shorthand notation
\begin{eqnarray*}
Q_j&=&\frac{1}{(j-2)^\delta + (j-3)\cdot 2^\delta +1}\\
&+&\frac{1}{(j-1)^\delta + (j-3)\cdot 2^\delta +1}
\end{eqnarray*}

When $\delta>2$, the probability to remain an open book is finite,
although it vanishes very rapidly when $\delta$ approaches to 2 from
above:
\begin{equation}
\mathcal{P}_\infty \sim \exp\!\left[-\frac{6}{\delta-2}\right]
\end{equation}

When $\delta\leq 2$, the exact open book structure will be certainly
destroyed at some moment. A sufficiently large network is thus not
an open book exactly, yet the deviation from this structure is
rather small. Consider for concreteness the range $3/2<\delta<2$
where the number of degree-$3$ nodes keeps growing while the number
of nodes of degrees $\geq 4$ remains finite. The degree sequence
reads
\begin{equation}
\label{deg-3}
(k_1,k_2,\underbrace{3,\ldots,3}_{N_3}\,,\underbrace{2,\ldots,2}_{N-N_3})
\end{equation}
where $k_1$ and $k_2$ are the highest degrees, and where we have not
displayed a finite number of other nodes whose degrees are different
from 2 and 3. The two highest degrees $k_1$ and $k_2$ are slightly
smaller than $N$. To determine $k_1$ and $k_2$  we first recall that
the sum of all degrees is twice the total number of links,
\begin{equation}
\label{L} \sum_{j=1}^N k_j = 2L
\end{equation}
Since $L=2N-4$ when $m=2$ and the initial sequence is $(2,1,1)$, we
use \eqref{deg-3} and re-write \eqref{L} as
\begin{equation}
k_1+k_2+3N_3+2(N-N_3) = 4N +O(1)
\end{equation}
{}from which $k_1+k_2 = 2N -N_3 + O(1)$. Combining this relation
with inequalities $k_1<N$ and $k_2<N$, we obtain
\begin{equation}
\label{kk} k_1 = N - pN_3, \quad k_2 = N - (1-p)N_3
\end{equation}
We now argue that $p=1/2$. Indeed, in the leading order the
difference $k_1-k_2$ evolves according to the rate equation
\begin{eqnarray}
\label{bias} \frac{d}{dN}\,(k_1-k_2) &=&
\frac{k_1^\delta-k_2^\delta}{2\cdot N^\delta}\,
\frac{N\cdot 2^\delta}{N^\delta}\nonumber\\
 &=& \frac{\delta\cdot 2^{\delta-1}}{N^\delta}\,(k_1-k_2)
\end{eqnarray}
This suggests that $k_1-k_2$ remains finite and therefore supports
\eqref{kk} with $p=1/2$. While the latter assertion is correct,
equation \eqref{bias} just shows that bias in favor of the node of
the highest degree $k_1$ over the second highest degree $k_2$ is too
small. However, there remain pure stochastic fluctuations, and the
difference $k_1-k_2$ is therefore a random variable of the order of
$\sqrt{N_3}$. Thus
\begin{equation}
\label{kk-stoch} k_1 = N - \frac{1}{2}\,N_3, \quad k_1-k_2 =
O(\sqrt{N_3})
\end{equation}

Let us now compute $N_3$. In the leading order we have
\begin{equation}
\label{N3-simple} \frac{d N_3}{d N} = 1-\mathcal{P}_{N\mapsto N+1}
\end{equation}
where for $\mathcal{P}_{N\mapsto N+1}$ we should ignore $O(N_3)$
corrections,
\begin{equation}
\label{PNN} \mathcal{P}_{N\mapsto N+1} =
\frac{N^\delta+N^\delta}{N^\delta+N^\delta+N\cdot 2^\delta}\cdot
\frac{N^\delta}{N^\delta+N\cdot 2^\delta}
\end{equation}
Plugging \eqref{PNN} into \eqref{N3-simple} and keeping only the
leading contribution we get
\begin{equation}
\frac{d N_3}{d N} = 3\left(\frac{2}{N}\right)^{\delta-1}
\end{equation}
which leads to $N_3 = a\,N^{2-\delta}$ from \eqref{N3-sol}.

To extract the sub-leading term, both \eqref{N3-simple}  and
\eqref{PNN} should be modified. To modify $\mathcal{P}_{N\mapsto
N+1}$ we use \eqref{deg-3} and \eqref{kk-stoch} and get
a more accurate formula for
\begin{eqnarray*}
\mathcal{P}_{N\mapsto N+1} & = &\frac{2 \cdot(N-N_3/2)^\delta}
{2 \cdot(N-N_3/2)^\delta+(N-N_3)2^\delta+N_3  \cdot 3^\delta}\\
&\times& \frac{(N-N_3/2)^\delta} {(N-N_3/2)^\delta+(N-N_3) \cdot
2^\delta+N_3 \cdot 3^\delta}
\end{eqnarray*}
The modification of \eqref{N3-simple} is
\begin{equation}
\label{N3-mod} \frac{d N_3}{d N} = 1-\mathcal{P}_{N\mapsto N+1} -
3\,\frac{N_3  \cdot 3^\delta} {N^\delta}
\end{equation}
where the last term on the right-hand side assures that whenever the
new node links to a node of degree 3, we have a loss rather than
gain. After lengthy calculations one gets
\begin{equation}
\label{N3_sol} N_3(N)\approx
\begin{cases}
a\,N^{2-\delta} + O(1) & \text{if $3/2<\delta<2$}\\
a\,N^{2-\delta} - b N^{3-2\delta}& \text{if $\delta<3/2$}\\
\end{cases}
\end{equation}
Strictly speaking, in writing  $\mathcal{P}_{N\mapsto N+1}$ we {\em
assumed} that $\delta>3/2$. However, a more detailed analysis shows
that the nodes of degree 4 do not influence the sub-leading
correction $b N^{3-2\delta}$.

When $4/3<\delta<3/2$, the nodes  of degree 4 become visible, and
the network degree sequence becomes
\begin{equation}
(k_1,k_2,\underbrace{4,\ldots,4}_{N_4}\,,\underbrace{3,\ldots,3}_{N_3}\,,\underbrace{2,\ldots,2}_{N-N_3-N_4})
\end{equation}
A straightforward generalization of our previous argument gives
\begin{equation}
\label{kk4} k_1 = N - \frac{1}{2}\,N_3 - N_4, \quad k_1-k_2 =
O(\sqrt{N_3})
\end{equation}
In the leading order, the quantity $N_4$ evolves according to
\begin{equation}
\frac{dN_4}{dN} = \frac{N_3 \cdot 3^{\delta}}{2\cdot N^\delta} +
\frac{N_3 \cdot 3^{\delta}}{N^\delta}
\end{equation}
{}from which
\begin{equation}
N_4(N) = a_4 N^{3-2\delta}\,,\quad a_4 =
\frac{3^{\delta+1}}{2}\,\frac{a}{3-2\delta}
\end{equation}

Proceeding the same way we obtain for any $k \geq 2$
\begin{equation}
\frac{dN_{k+1}}{dN} = \frac{3}{2}\,\frac{N_k \cdot
k^{\delta}}{N^\delta}
\end{equation}
leading to the asymptotic
\begin{equation}
N_{k+1}(N) = a_{k+1}N^{k-(k-1)\delta}
\end{equation}
with amplitudes
\begin{equation}
a_{k+1} = a\left(\frac{3}{2}\right)^{k-2}\prod_{j=3}^k
\frac{j^\delta}{j-(j-1)\delta}
\end{equation}

\end{document}